\documentclass[runningheads,a4paper]{llncs}

\usepackage{amssymb}
\usepackage{graphicx}
\usepackage{amsfonts}
\usepackage{calligra}
\usepackage{amsmath}
\usepackage{subfigure}
\usepackage{multirow}
\usepackage{siunitx, booktabs}
\usepackage{makecell}
\usepackage{bm}
\usepackage{breqn}
\usepackage{amsmath}

\DeclareMathOperator*{\argmin}{arg\,min}

\usepackage{url}

\begin{document}

\mainmatter  

\title{Generative Adversarial Networks (GAN) Powered Fast Magnetic Resonance Imaging\\---~Mini Review, Comparison and Perspectives}

\titlerunning{GAN Powered Fast MRI}

%
%
\author{Guang Yang\inst{1}$^{,}$\inst{2}%
\thanks{Corresponding author: g.yang@imperial.ac.uk.}%
\and Jun Lv\inst{3} \and Yutong Chen\inst{1}$^{,}$\inst{4} \and Jiahao Huang\inst{5} \and Jin Zhu\inst{6}}
\authorrunning{Yang et al.}

\institute{National Heart \& Lung Institute, Imperial College London, London SW7 2AZ, UK \and
Cardiovascular Research Centre, Royal Brompton Hospital, London, SW3 6NP, UK \and
School of Computer and Control Engineering, Yantai University, Yantai, China \and
Department of Physiology, Development and Neuroscience, University of Cambridge, CB2 3EG, UK \and
School of Optics and Photonics, Beijing Institute of Technology, Beijing, China \and
Department of Computer Science and Technology, University of Cambridge, Cambridge, CB3 0FD, UK}

%
%

\maketitle

\begin{abstract}

Magnetic Resonance Imaging (MRI) is a vital component of medical imaging. When compared to other image modalities, it has advantages such as the absence of radiation, superior soft tissue contrast, and complementary multiple sequence information. However, one drawback of MRI is its comparatively slow scanning and reconstruction compared to other image modalities, limiting its usage in some clinical applications when imaging time is critical. Traditional compressive sensing based MRI (CS-MRI) reconstruction can speed up MRI acquisition, but suffers from a long iterative process and noise-induced artefacts. Recently, Deep Neural Networks (DNNs) have been used in sparse MRI reconstruction models to recreate relatively high-quality images from heavily undersampled \textit{k}-space data, allowing for much faster MRI scanning. However, there are still some hurdles to tackle. For example, directly training DNNs based on L1/L2 distance to the target fully sampled images could result in blurry reconstruction because L1/L2 loss can only enforce overall image or patch similarity and does not take into account local information such as anatomical sharpness. It is also hard to preserve fine image details while maintaining a natural appearance. More recently, Generative Adversarial Networks (GAN) based methods are proposed to solve fast MRI with enhanced image perceptual quality. The encoder obtains a latent space for the undersampling image, and the image is reconstructed by the decoder using the GAN loss. In this chapter, we review the GAN powered fast MRI methods with a comparative study on various anatomical datasets to demonstrate the generalisability and robustness of this kind of fast MRI while providing future perspectives. 

\end{abstract}

\section{Introduction}

\subsection{Magnetic Resonance Imaging}

Through offering repeatable, non-invasive measures of tissue structure and function, Magnetic resonance imaging (MRI) has transformed clinical medical imaging and medical science. The sensitivity of the image to tissue properties can be greatly varied with MRI, either by changing the timing with which MR signals are obtained (e.g., the echo time---TE and the repetition time---TR), or by using magnetisation preparation or contrast agents. These so-called multimodal or multi-sequence MRI methods can not only provide a comparison in traditional anatomical or structural MRI but can also quantify the function of most tissues and organs of the human body in clinical and pre-clinical laboratory environments. Invasive methods, such as tissue biopsy or radionuclide tests, have therefore become less needed as a result of the prosperity of MRI and other non-invasive medical imaging technologies \cite{Hollingsworth2015}. Since MRI is non-invasive, it can be used to provide longitudinal and quantitative imaging biomarkers in the therapy trials. MR methods have begun to represent gold standard measurement for clinical research, despite the fact that the modality is still relatively underutilised \cite{Hollingsworth2015}. Why is this so? 'MRI is complex and pricey,' which is a commonly heard melancholy.

\subsection{Limitations of Magnetic Resonance Imaging}

Although MRI is a revolutionary non-invasive diagnostic imaging technique that provides high-resolution definition of the structural and functional information of most body tissues and organs, one significant drawback of MRI is its slow rate of image acquisition, which results in a longer scanning period as compared to other imaging modalities \cite{Lustig2007,Yang2018}.

In MRI, raw data is acquired in the \textit{k}-space, which includes information about spatial frequencies within the image, rather than collected directly in the image space \cite{Suetens2009}. The Fourier transformation links image space and the \textit{k}-space. The Nyquist criterion defines the \textit{k}-space information that must be satisfied conventionally after we have defined the field-of-view (FOV) and spatial resolution of the image that we want to obtain. The distance between \textit{k}-space neighbours is inversely proportional to the field of view in each direction. The highest frequency obtained in each direction is inversely proportional to the desired resolution. Data encoded with pulsed magnetic field gradients are acquired to fill the \textit{k}-space. We may obtain a line of \textit{k}-space points very quickly in one direction, known as the read direction, using either a spin or gradient echo in one repetition time. Further directions, on the other hand, must be phase-encoded, which takes one repetition time to encrypt one line of \textit{k}-space lines \cite{Hollingsworth2015,Suetens2009}. This must then be rerun for all possible combinations of the number of phase encoding steps needed in the anterior-posterior and foot-head directions. As a result, MRI acquisitions can be time-consuming, particularly when a high resolution or large FOV is needed.

This drawback not only raises the cost of imaging but also limits its use in emergency situations. Furthermore, in order to maintain image consistency, patients must lie still during the acquisition. For abdominal/thoracic imaging, patients must hold their breath, which can be problematic for paediatric, obese patients, and those with respiratory compromise \cite{Hollingsworth2015}. As a result, many patients can experience anxiety and claustrophobia during the MR scanning procedure \cite{Hollingsworth2015}. In order to minimise scanning costs and increase patient throughput, the MR image acquisition process must be sped up. Various MR acceleration methods rely on taking measurements of several lines per repetition time, allowing for quicker traversing of the \textit{k}-space. Examples include echo planar imaging \cite{Mansfield1977}, rapid acquisition with relaxation enhancement \cite{Hennig1986}, and fast low angle shot imaging \cite{Haase1986}.

\subsection{Conventional Acceleration Using Compressive Sensing}

It is possible to attain a higher degree of acceleration by sampling the \textit{k}-space only partially, i.e., not collecting all lines of measurements in the phase encoding path(s). The undersampled calculation can be used to infer the original \textit{k}-space details. As a consequence, the acceleration metric is equal to the undersampling ratio. For example, if half of the \textit{k}-space is sampled, the acceleration factor is doubled. As a result, undersampling methods aim to circumvent the Nyquist-Shannon sampling criterion \cite{Lustig2007}.

Compressed sensing (CS) is a promising undersampling approach that may allow for more aggressive undersampling and acceleration \cite{Lustig2007}. CS principle is similar to the concept of compressing signals for transmission and then decompressing them \cite{Zisselman2018}, as seen in the JPEG, JPEG2000, MPEG, and MP3 standards \cite{Hollingsworth2015}. If undersampled signals or images can be compressed correctly, they can also be decompressed or recovered accurately, according to CS \cite{Fair2015}. Hereby, CS sets three conditions on the MRI reconstruction: 

\begin{enumerate}
    \item The image or signal must be compressible. In other words, the MRI images must be sparse, with the bulk of its pixel values being zeros, either in its native domain or in an appropriate transformation domain, such as in the wavelet or frequency domain. 
    \item To avoid aliasing artefacts, the undersampling patterns should be incoherent using random undersampling. 
    \item A non-linear reconstruction algorithm must be used. 
\end{enumerate}

It is possible to recover the original MRI images from their undersampled measurements following these three criteria.

Previously published research on using CS as an MR acceleration approach employs iterative non-linear optimisation algorithms that implement sparsity and reconstruction fidelity. Total variation (TV) \cite{Lustig2007}, dictionary learning (DLMRI \cite{Ravishankar2011}, RecPF \cite{Yang2010}, and BM3D \cite{Eksioglu2016}) are typical examples. However, there are four major issues with these approaches:
\begin{enumerate}
    \item Iterative optimisation can be time-consuming \cite{Hollingsworth2015,Hu2014}.
    \item These algorithms tend to generate artificially smoothed image appearance \cite{Hollingsworth2015}.
    \item The reconstruction results can have blocky artefacts \cite{Liu2015,Kayvanrad2014,Guerquin-Kern2011}.
    \item They reconstruct each image as an individual event, failing to account for the expected anatomical features in MR images that may be used to improve the reconstruction accuracy \cite{Hammernik2018}.
\end{enumerate}

\subsection{Deep Learning Based Fast MRI}

Deep learning based approaches have recently achieved performance dividends in a variety of medical image processing problems by using ‘big data' and advancements in computational power. To date, however, the majority of research studies have concentrated on downstream medical image interpretation and post-processing activities, such as anatomical segmentation \cite{chen2021jas,wu2021fast,wu2021automated,jin20213d,zhou2020systematic,liu2020exploring,ferreira2020automating,li2020mv,liu2019automatic,zhuang2019evaluation,mo2018deep}, lesion segmentation \cite{zhang2021me,yang2020simultaneous,li2020atrial,zhang2019automatic,yang20186,bakas2018identifying}, co-registration \cite{mok2020fast,de2017end,wu2015scalable}, synthesis \cite{xu2021synthesis,wang2021dicyc}, and multimodal data detection \cite{gao2020salient,wang2019saliencygan,ali2020novel,yang2020deep,li2018deep,dong2018holistic,dong2017automatic}, for disease identification \cite{hu2020weakly,cao2020multiparameter,zhang2019deep}, prognosis \cite{roberts2020machine,soltaninejad2017mri}, and treatment prediction \cite{jin2021predicting,nielsen2018prediction}. To increase the precision of these post-processing operations, imaging methods must be improved, which can also be aided by deep learning \cite{chen2021wavelet,lv2021pic,lvgan,yuan2020sara,guo2020deep,schlemper2018stochastic,seitzer2018adversarial}.

Since its principle was developed in 2006, CS has had a long history for fast imaging applications, including the embodiment of MRI reconstruction \cite{donoho2006compressed}. However, the related less efficient iterative optimisation can stymie further implementation. Although deep learning based tomographic reconstruction technology has only been around for a few years, there is a lot of interest in this area, and there are many ongoing advances and exciting applications, including MRI.

Deep learning based approaches can successfully overcome the majority of the aforementioned shortcomings of earlier CS methods. A deep learning algorithm, e.g., convolutional neural networks (CNN), is made up of many layers of nodes. To learn the mapping from undersampled MR images to their corresponding fully sampled ones, the weights of the node relations between layers are optimised. The method of optimising weights is known as training the model. Once trained, the model is capable of reconstructing original images from undersampled measurements. In terms of reconstruction accuracy, speed, and visual consistency, deep learning based methods have been shown to consistently outperform non-deep learning based ones \cite{Quan2018,Mardani2019,Yang2018,Schlemper2018,Huang2019,Eo2018}.

\subsection{GAN Powered Fast MRI}

Generative Adversarial Networks, or GAN for short, represent a type of generative modelling technique that employs deep learning methods, e.g., CNN. Generative modelling is an unsupervised learning task in machine learning that entails automatically finding and learning the regularities or patterns in input data such that the model can be used to produce or output new examples that could have been plausibly drawn from the original dataset.

GAN employs a clever method for training a generative model by posing the problem as a supervised learning problem with two sub-models: the generator model, which we train to produce new examples, and the discriminator model, which attempts to identify examples as either true (from the original domain) or false (generated). The two models are trained in an adversarial zero-sum game before the discriminator model is tricked about half of the time, indicating that the generator model is producing plausible instances. 

GAN is a fascinating and quickly evolving area that delivers on the promise of generative models by producing plausible instances in a variety of problems, most notably in image-to-image conversion tasks such as translating image styles, and in generating photo-realistic images of objects, scenes, and individuals that even humans cannot recognise which ones are fake.

GAN is an important type of deep learning based CS-MRI reconstruction method, which was proposed first by Yang et al. in 2017 \cite{yu2017deep,Yang2018}. In the context of CS-MRI, GAN entails training a generator to recreate the original image from undersampled measurements and a discriminator to produce the likelihood of whether the generated image matches the original, i.e., fully sampled measurements. The discriminator, in turn, modifies the generator's learning \cite{Goodfellow2014}. As a consequence, the generator generates photo-realistic images \cite{Deng2019}. In terms of reconstruction accuracy and efficiency, GAN based methods \cite{Yang2018,Quan2018} outperform the non-GAN based deep learning method, e.g., deep ADMM-net. One GAN based approach \cite{Mardani2019} also claims to generate less fuzzy and aliasing artefacts than non-deep learning based methods. As a result, GAN based approaches have the capability to produce state-of-the-art CS-MRI reconstruction results.

~\\
In this book chapter,
\begin{itemize}
    \item we will perform a mini topical review on GAN powered fast MRI, including the original Deep De-Aliasing Generative Adversarial Networks (DAGAN) method and other more advanced and recently proposed GAN based models;
    \item we will analyse and explain different GAN models, and compare the results obtained by different GAN based models;
    \item we will provide a comparison study on different datasets, e.g., MRI for various anatomical scans.
    \item we will highlight the recent development and discuss future directions.
\end{itemize}

\section{Methods}

\subsection{Fundamentals of MRI Reconstruction}
\sloppy

The basis of undersampled MR reconstruction is inferring the missing \textit{k}-space data from the already sampled \textit{k}-space values. For example, in partial Fourier imaging (PFI), 50\% of the \textit{k}-space is acquired. Because the \textit{k}-space is conjugate symmetric, the other missing 50\% is obtained via complex conjugation of the existing 50\%. However, the maximum acceleration is only 2-fold and the reconstructed image exhibits a lower signal-to-noise ratio (SNR) \cite{Hollingsworth2015}.  Another key undersampling technique is parallel imaging (PI). In PI, multiple receiver coils simultaneously collect the \textit{k}-space information of the tissues closest to each coil. However, the acceleration factor is limited by the number and the configuration of the receiver coils \cite{Hollingsworth2015,deshmane_parallel_2012}. Therefore, limits of accelerating MR image acquisition exist for both PI and PFI, bound by the Nyquist-Shannon sampling criteria \cite{Lustig2007}.

Compressed sensing (CS)-based reconstruction circumvents the Nyquist-Shannon sampling criteria and achieves a higher acceleration ratio. CS establishes the model of MR image acquisition as (notations in Table \ref{tab:CS_notation}):
\begin {equation}
    y_u = Ax_t,
\label{eq:CS_acquisition}\end {equation}
where $y_u$ is the undersampled \textit{k}-space signal and $x_t$ is the original fully sampled image. $A$ is an operator defined as: 
\begin{equation}
    A = \Psi \mathcal{F},
\end{equation}
where $\Psi$ is the undersampling mask, a binary matrix denoting which \textit{k}-space locations are sampled and which are not, and $\mathcal{F}$ is the Fourier transform operator. Hence, $\Psi \mathcal{F} x_t$, or $Ax_t$ is equivalent to undersampling the
\textit{k}-space of the putative reconstructed MR image. 

A CS-model infers the underlying full resolution MR image $x_t$ from the collected incomplete \textit{k}-space samples $y_u$ by solving the following optimisation problem \cite{Lustig2007}:
\begin{equation}
    \argmin_{\hat x_u} \frac{\lambda}{2}||y_u-A \hat x_u||_2^2 + R(\hat x_u),
\label{eq:CS_eq}\end{equation}
where $\lambda$ adjusts the contribution of the first term to the optimisation objective and $R(\hat x_u)$ is an image regulariser function.  If this reconstructed image $\hat x_u$ is to match the original image, their undersampled \textit{k}-space results ought to match, as reflected in the first term of Eq \ref{eq:CS_eq}. This reinforces the \textit{k}-space data fidelity of reconstructing the undersampled data.

The second term in Eq \ref{eq:CS_eq} ensures the reconstructed image possesses certain attributes such as smoothness or sparsity, which is required by the theory of CS. 
For example, total variation---an early CS technique---uses the following regulariser term to ensure the underlying image is smooth \cite{Lustig2007,zhang_deep_2020-2}:
\begin{equation}
    \argmin_{\hat x_u} \frac{\lambda}{2}||y_u-A \hat x_u||_2^2 + ||\nabla \hat x_u||_1,
\label{eq:CS_TV}\end{equation}
where $\nabla$ is the gradient operator to minimise the difference between adjacent pixels in the final reconstruction to ensure a smooth texture. To solve Eq \ref{eq:CS_TV}, conjugate gradient descent is employed \cite{Lustig2007}, by updating the reconstructed image with the gradient of Eq \ref{eq:CS_TV} iteratively until convergence.

Another example of CS algorithm is dictionary learning, in which the regulariser function enforces the sparsity of the dictionary representation of each image patch \cite{Ravishankar2011}:
\begin{equation}
    \argmin_{\hat x_u, D, \Gamma} \frac{\lambda}{2}||y_u-A \hat x_u||_2^2 + \sum_{ij}||R_{ij} \hat x_u -Da_{ij}||_2^2,
\label{eq:CS_DL}\end{equation}
where $R_{ij}$ is the image patch extractor, $D$ is the dictionary used to transform the dictionary representation $a_{ij}$ into the image domain, and $\Gamma$ is the set of all dictionary representations $\{a_{ij}\}_{ij}$. This optimisation problem in Eq \ref{eq:CS_DL} is solved by alternating minimisation of the 3 parameter sets iteratively: $\hat x_u$, $D$ and $\Gamma$ collectively. CS algorithms such as TV and dictionary learning repeatedly update the reconstruction to ensure it fulfils \textit{k}-space data fidelity and certain regularisation properties such as sparsity.

From a Bayesian perspective, data fidelity is seen as maximising the conditional probability of observing the undersampled \textit{k}-space data given the reconstructed image. The regulariser term represents the prior expectation of the statistical properties of the final reconstructed image \cite{diamond_unrolled_2017}. By jointly enforcing data fidelity and regulariser constraint, CS reconstruction is equivalent to maximising the posterior probability that the reconstructed image matches the original one given the existing \textit{k}-space samples. This justifies the effectiveness of the CS in reconstructing undersampled MR
images.

\begin{table}
\begin{center}
\begin{tabular}{ll}
        \toprule
        \textbf{Symbol} & \textbf{Definition} \\
        \midrule
        $x_t$ & Ground truth MR image \\
        $x_u$ & Undersampled MR image \\
        $\hat x_u$ & Reconstructed MR image \\
        $y_t$ & Ground truth \textit{k}-space signals \\
        $y_u$ & Undersampled \textit{k}-space signals \\
        $\mathcal{F}$ & Fourier transform operator \\
        $\Psi$ & Binary matrix denoting undersampled \textit{k}-space positions  \\
        $A$ & $\Psi \mathcal{F}$  \\
        $R$ & Regularisation function \\
        $\lambda$ & \makecell[l]{ A parameter adjusting the contribution of data \\ fidelity term to the reconstruction loss function} \\
        $\theta$ & Parameters within a convolutional neural network \\
        \bottomrule
\end{tabular}
\end{center}
\caption{Mathematical notations in this book chapter.}
\label{tab:CS_notation}\end{table}

\subsection{CNN Based MRI Reconstruction}
CNN excels in computer vision applications compared to traditional machine learning methods in general, and has been increasingly incorporated into CS-based MRI reconstruction models. It achieves higher accuracy and acceleration of MRI acquisition compared to traditional CS models \cite{Yang2018,Eo2018,Schlemper2018,qin_convolutional_2019}. A CNN consists of a series of convolutional layers connected by non-linear activation functions \cite{jogin_feature_2018}. In each convolutional layer, multiple filters are convolved with the input image or the output from a previous layer to extract image specific features \cite{jogin_feature_2018}. The weight of each filter is optimised such that the final reconstructed image matches the original fully sampled image. By stacking multiple convolutional layers, the large number of weight parameters in a CNN endows it with the potential to model complex functions, including recovering the fully sampled images from the undersampled ones.

The process of optimising the weights of CNN is known as training the network and is governed by an optimisation objective called the loss function. The loss functions differ between supervised and unsupervised settings, depending on whether fully sampled ground truth images are available. In a supervised learning setting, fully sampled images are used to ``teach'' the CNN model to minimise the loss, or difference, between the reconstructed results and the fully sampled ground truth. However, in an unsupervised environment, ground truth images are not available, meaning the loss function can be, for example, optimised based on the Deep Image Prior framework and uses a high-resolution reference MR image as the input of the CNN to induce the structural prior in the learning procedure \cite{zhao_reference-driven_2020}. Unsupervised CNN-based CS-MRI models are beyond the scope of this chapter, which will instead focus on the supervised learning models.

Supervised CNN-based CS-MRI methods can be broadly divided into two categories: end-to-end and unrolled optimisation \cite{liang_deep_2019,liang_deep_2020,zhang_review_2020}.  End-to-end methods model the inverse of the acquisition process (Eq \ref{eq:CS_acquisition}) by mapping the undersampled input to the reconstructed output directly, hence the name ``end-to-end'':
\begin{equation}
    \hat x_u = f_{\mathrm{CNN}} (y_u|\theta),
\end{equation}
where $f_{CNN}$ is the operation performed by the CNN and $\theta$ represents all the parameters within the CNN. The $\theta$ is optimised by minimising the difference between $\hat x_u$ and the ground truth $x_t$.  Such end-to-end techniques are exemplified by U-Net \cite{lee_deep_2018} and generative adversarial network (GAN) \cite{Yang2018,Mardani2019,Quan2018}. 

In contrast, CNN-based unrolled optimisation methods perform iterative image update based upon the general CS reconstruction model (Eq \ref{eq:CS_eq}). Unlike non-CNN based CS techniques, where the term $R(\hat x_u)$ is an expert-designed regularisation function, the CNN-based unrolled methods apply CNN to learn the optimal way of regularising an image. This is exemplified by the deep cascade CNN (DC-CNN) network \cite{Schlemper2018}, whose regulariser term penalises the deviation of the reconstructed image from the CNN output:
\begin{equation}
    \argmin_{\hat x_u, \theta} \frac{\lambda}{2}||y_u-A \hat x_u||_2^2 + ||\hat x_u-f_{\mathrm{CNN}}(y_u|\theta)||.
\label{eq:CNN_DCCNN}\end{equation}

Another example is the variational network \cite{Hammernik2018}, which uses a Field-of-Expert regularisation function:

\begin{equation}
    \argmin_{\hat x_u, \theta} \frac{\lambda}{2}||y_u-A\hat x_u||_2^2 + \sum_i f_i(k_i\hat x_u),
\label{eq:CNN_VN}\end{equation}
where $f_i$ is a learnable activation function and $k_i$ represents a convolutional filter. From a Bayesian point of view, these CNN-based regularisers form the prior expectation of the reconstructed images by learning from the ground truth MR
images.

While both end-to-end and unrolled methods can incorporate CNN to learn the image reconstruction or regularisation process, a key difference between them is that unrolled methods need to iteratively update the images during the reconstruction process. In contrast, end-to-end ones compute the output directly. 
Hence, it is more time consuming to train unrolled CNN-based models and to apply them in reconstruction, compared with end-to-end methods, with other comparisons summarised in Table \ref{tab:UO_ETE}.  This suggests an advantage of developing end-to-end methods, such as GAN-based models, for CS-MRI reconstruction.

\begin{table}
\begin{center}
\begin{tabular}{lll}
        \toprule
        \textbf{Category} & \textbf{End-to-end} & \textbf{Unrolled} \\
        \midrule
        Reconstruction time & Short & Long  \\
        Data fidelity & No    & Yes  \\
        Sample size   & Larger & Smaller\\
        Performance   & Lower & Higher\\
        Weight update & Yes   & Yes  \\
        Image update  & No    & Yes  \\
        Parameter number & Larger & Smaller \\
        \bottomrule
\end{tabular}
\end{center}
\caption{Comparison between end-to-end and unrolled optimisation methods in
        CNN-based CS-MRI models \cite{liang_deep_2019,liang_deep_2020,zhang_review_2020}.}
\label{tab:UO_ETE}\end{table}

\subsection{GAN Based MRI Reconstruction}
\subsubsection{General GAN}

Inspired by the two-player zero-sum game in game theory, GAN \cite{Goodfellow2014} consists of two players: a generator and a discriminator. Traditionally, the generator $G$ captures the distribution of sample data, and uses noise $z$ that follows a certain distribution to generate a sample $G(z)$ similar to the real training data $x_t$. The discriminator $D$ is a binary classifier, which aims to distinguish fake data generated by $G$ from the ground truth. If the sample comes from the real training data, $D$ outputs a large probability, otherwise, $D$ outputs a small probability. Ideally, the best $D$ can be represented as $D(x_t)=1$ and $D(G(z))=0$. In this way, the generator and the discriminator form a min-max game. The training process of GAN can be described as follows:

\begin{dmath}
\mathop{\text{min}}\limits_{\theta_G} \mathop{\text{max}}\limits_{\theta_D}
L(\theta_G, \theta_D)
=\mathbb{E}_{\bm{x} \sim p_{\mathrm{data}}(\bm{x})}
[\mathop{\text{log}} D_{\theta_D}(\bm{x})] ~\\
+\mathbb{E}_{\bm{z} \sim p_{\bm{z}}(\bm{z})}
[\mathop{\text{log}} (1-D_{\theta_D}(G_{\theta_G}(\bm{z}))],
\end{dmath}
where $p_{\mathrm{data}}(\bm{x})$ is the distribution of the training dataset, and the $p_{\bm{z}}(\bm{z})$ is the distribution of the latent variables. 

Alternating gradient optimisation between $G$ and $D$ is used to train the GAN. In this process, both models try their best to optimise their networks to form a competitive confrontation until the two models reach a dynamic balance. Ideally, the final result is that the image generated by $G$ is very similar to the real image, and it is difficult for the $D$ network to distinguish between the real image and the image generated by $G$, i.e., $D(G(Z)) = 0.5$.

\subsubsection{DAGAN}

\begin{figure}[!h]
\centering\includegraphics[width=5in]{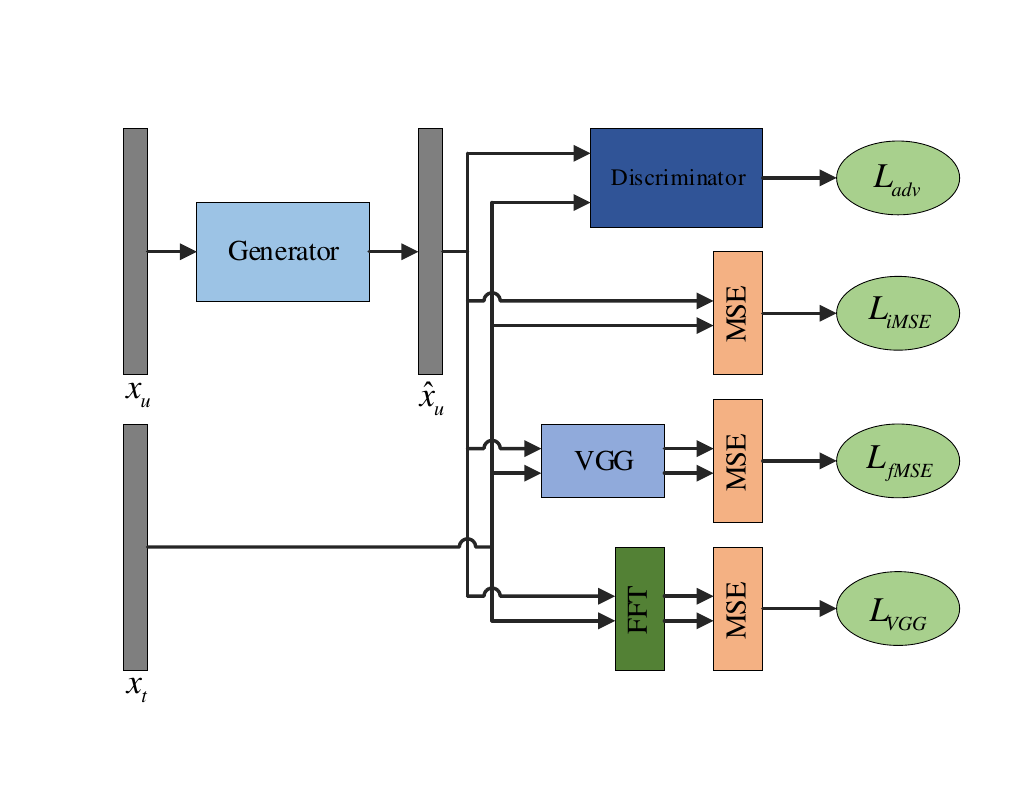}
\caption{The architecture of DAGAN. The generator produced the MR image $\hat x_u$ from the undersampled MR image. The generated MR image $\hat x_u$ and the ground truth MR image were input into the discriminator for adversarial loss $L_{\mathrm{adv}}$ calculation. The data consistency loss consisted of image domain MSE loss $L_{\mathrm{iMSE}}$, frequency domain MSE loss $L_{\mathrm{fMSE}}$, and VGG perceptual loss $L_{\mathrm{VGG}}$.}
\label{dagan}
\end{figure}

Deep De-Aliasing Generative Adversarial Networks (DAGAN) was proposed in 2017 by Yang et al. \cite{Yang2018,yu2017deep} for fast compressed sensing MRI reconstruction. Figure \ref{dagan} shows the architecture of DAGAN.

\begin{figure}[!h]
\centering\includegraphics[width=5in]{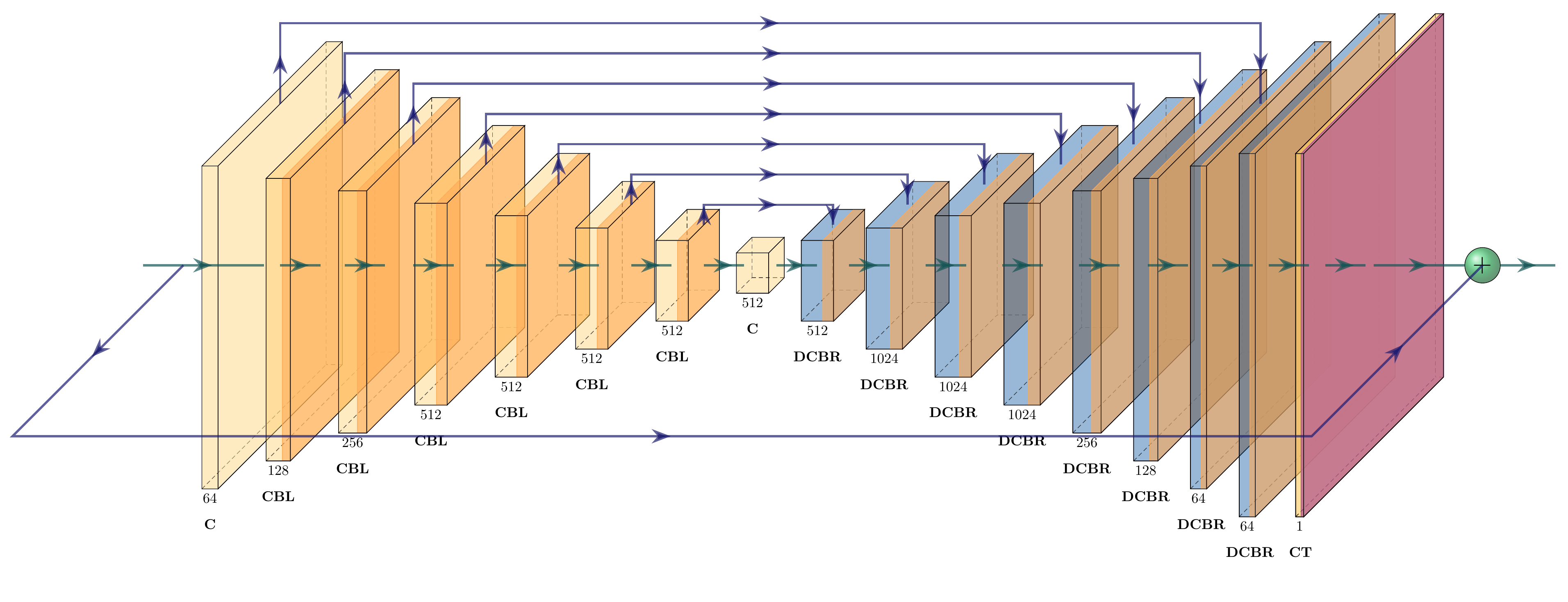}
\caption{Structure of the generator in DAGAN. The generator was based on a modified U-Net, which consisted of 8 convolutional layers and 8 transposed convolutional layers (de-convolutional layers), with a BN layer and a LReLU or ReLu layer followed with each layer. Skip connection between layers that were the same scale and shortcut connections between the input and output of the generator were applied. A hyperbolic tangent function was used as an output activation function. (C: Convolutional layer; DC: De-Convolutional layer; B: Batch Normalization layer; L: Leaky ReLU layer; R: ReLU layer; T: hyperbolic tangent function.)}
\label{dagan_generator}
\end{figure}

As shown in Figure \ref{dagan_generator}, a modified U-Net \cite{Ronneberger2015} was used as the generator $G$, which consisted of 8 convolutional layers in the encoding path and 8 deconvolutional layers in the decoding path. The stride of all convolutional and deconvolutional layers was set to 2 for downsampling and upsampling the feature maps. Each convolutional layer was followed by a Batch Normalisation (BN) layer and a Leaky ReLU (LReLU) layer. Skip connection was applied between corresponding layers in the encoding and decoding paths in order to pass the features in the encoding layer to the decoding layer for better reconstruction details. The modified U-Net ended up with a hyperbolic tangent function as the activation function. Due to the alternating training strategy in the adversarial components, the original GAN model is hard to train. DAGAN applied $\hat x_u=G(x_u)+x_u$ instead of $\hat x_u=G(x_u)$ as the output of the generator for better stability and faster convergence speed. In this way, DAGAN turned the generator into a refinement function, which means that it only generated the missing information, and the complexity of the model was reduced.

\begin{figure}[!h]
\centering\includegraphics[width=5in]{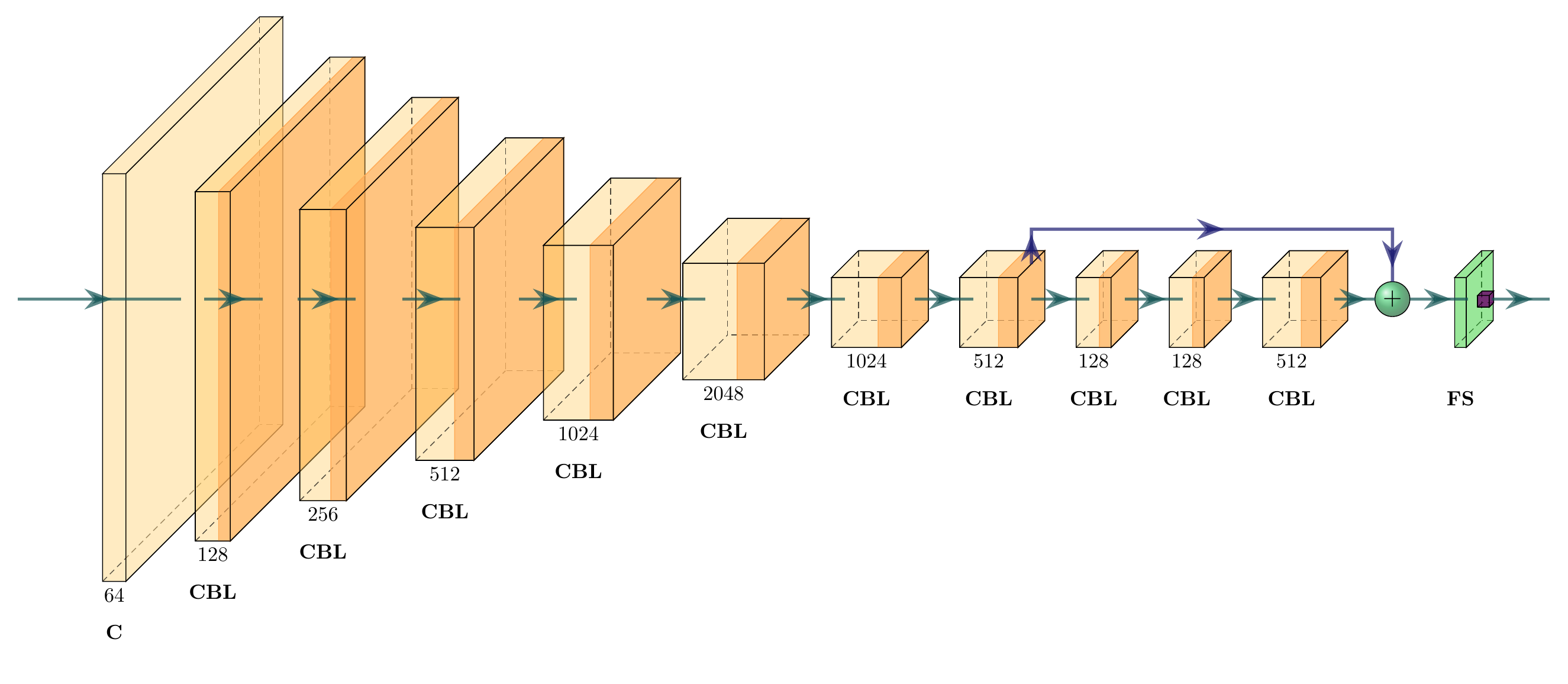}
\caption{Structure of the discriminator in DAGAN. Essentially, the discriminator was an 11-layer CNN classifier. Each convolutional layer was cascaded by a BN layer and a LReLu Layer. A full connection layer and a Sigmoid function were applied to output the result of the classification. (C: Convolutional layer; B: Batch Normalization layer; L: Leaky ReLU layer; F: Full Connection layer; S: Sigmoid function.)}
\label{dagan_discriminator}
\end{figure}

As shown in Figure \ref{dagan_discriminator}, an 11-layer CNN architecture was used as the discriminator $D$. A BN layer and a LReLU layer were followed with each convolutional layer. Finally, a full connection (FC) layer was cascaded, and the classification result was output through the Sigmoid activation function. 

In DAGAN, normalised MSE was used as the optimisation cost function. However, only optimising the pixel-wise MSE loss in the image domain could result in non-smooth reconstructions, which might lack coherent image details. To solve this problem, a data consistency loss was designed for training the generator in both frequency and image domains to help the optimisation and to exploit the complementary properties of the two domains. Besides, perceptual similarity \cite{Ledig_2017_CVPR} was also incorporated. In so doing, the data consistency loss consists of three parts: pixel-wise image domain MSE loss $L_{\mathrm{iMSE}}$, frequency domain MSE loss $L_{\mathrm{fMSE}}$, and the pre-trained VGG perceptual loss $L_{\mathrm{VGG}}$ \cite{simonyan2015deep}. They can be defined as:

\begin{equation}
L_{\mathrm{iMSE}}(\theta_G) =
\frac{1}{2}\mid\mid x_t -  \hat x_u \mid\mid^2_2,
\end{equation}

\begin{equation}
L_{\mathrm{fMSE}}(\theta_G) =
\frac{1}{2}\mid\mid \mathcal{F} x_t- \mathcal{F} \hat x_u \mid\mid^2_2,
\end{equation}

\begin{equation}
L_{\mathrm{VGG}}(\theta_G) =
\frac{1}{2}\mid\mid f_{\mathrm{VGG}}(x_t) - f_{\mathrm{VGG}}(\hat x_u) \mid\mid^2_2.
\end{equation}

Here, $\mid\mid\cdot\mid\mid^2_2$ denotes the L2 norm, $f_{\mathrm{VGG}}(\cdot)$ denotes VGG network (the Conv4 output of the VGG16), and was pretrained on the ImageNet \cite{RussakovskyDeng-363}.

The adversarial loss $L$ can be defined as:
\begin{dmath}
L_{\mathrm{adv}}(\theta_G, \theta_D)
=\mathbb{E}_{{x_t} \sim P_{\mathrm{train}}(x_t)}
[\mathop{\text{log}} D_{\theta_D}({x_t})] ~\\
+\mathbb{E}_{{x_u} \sim p_{G}(x_u)}
[\mathop{\text{log}} (1-D_{\theta_D}(G_{\theta_G}(x_u))],
\end{dmath}
where $P_{\mathrm{train}}(x_t)$ denotes the collection of MR image ground truth, and $p_{G}(x_u)$ denotes the collection of MR image reconstruction.

Therefore the total loss function can be represented as:
\begin{dmath}
L_{\mathrm{TOTAL}}(\theta_G, \theta_D)
= \alpha L_{\mathrm{iMSE}}(\theta_G)
+ \beta L_{\mathrm{fMSE}}(\theta_G)
+ \gamma L_{\mathrm{VGG}}(\theta_G) ~\\
+ L_{\mathrm{adv}}(\theta_G, \theta_D),
\end{dmath}
where $\alpha$, $\beta$ and $\gamma$ were the weights of different components of the loss function, which balanced different loss terms into similar scales according to previous study. Here, $\alpha=15$, $\beta=0.1$ and $\gamma=0.0025$ were set empirically, according to the original paper.

\subsubsection{KIGAN}

\begin{figure}[!h]
\centering\includegraphics[width=5in]{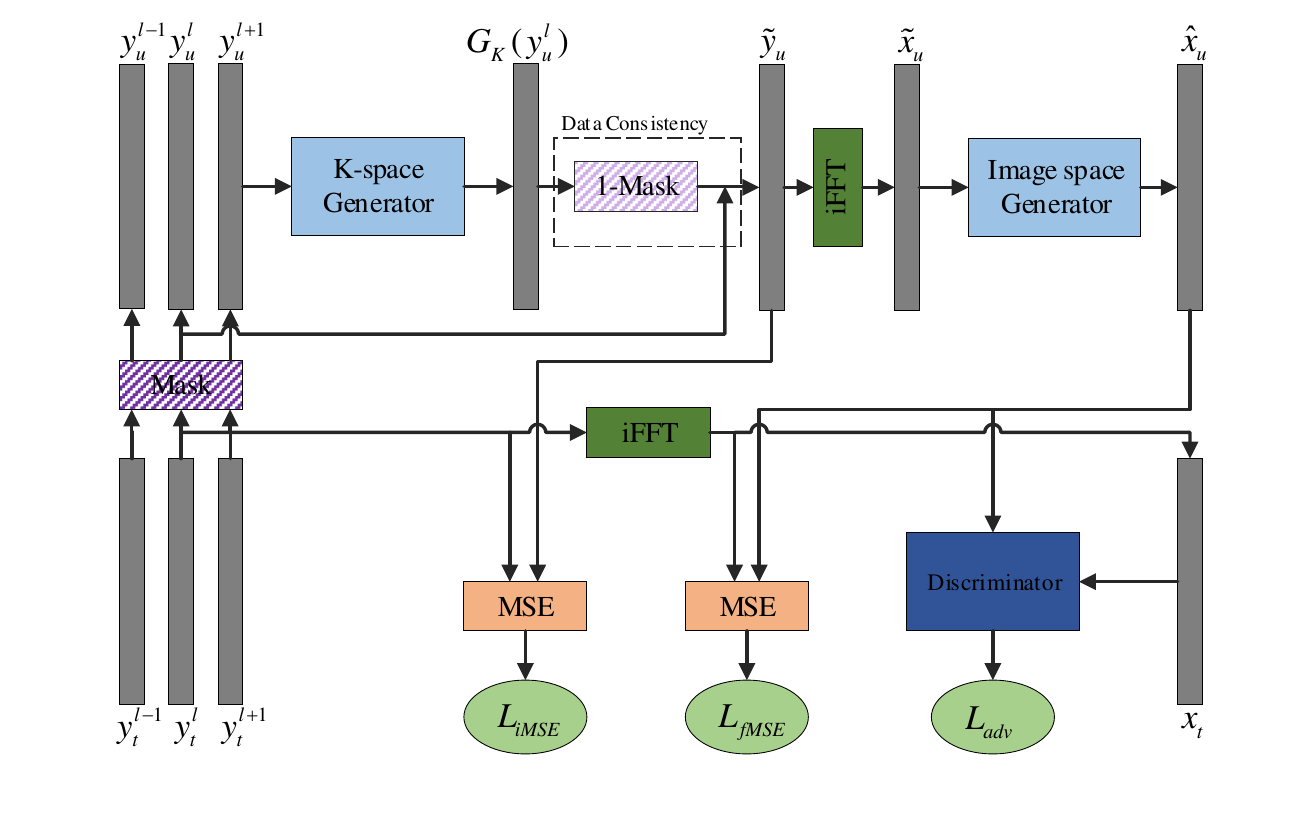}
\caption{The Architecture of KIGAN. Two generators and one discriminator were used in KIGAN. \textit{k}-space generator was able to reconstruct \textit{k}-space MR data $G_{\mathrm{K}}(y^{l}_u)$ from a cascade of undersampled \textit{k}-space MR data $y^{l-1}_u$, $y^{l}_u$ and $y^{l+1}_u$, and image space generator was able to reconstruct the MR image $\hat x_u$ (final result). The discriminator was to output the classification result of ground truth MR image $x_t$ and generated MR image $\hat x_u$ for adversarial loss $L_{\mathrm{loss}}$. Image domain MSE loss $L_{\mathrm{iMSE}}$ and frequency domain MSE loss were added into the total loss function.}
\label{kigan}
\end{figure}

KIGAN was introduced by Shaul et al. \cite{SHAUL2020101747}. The overall structure of KIGAN is shown in Figure \ref{kigan}. A \textit{k}-space generator $G_{\mathrm{K}}$ and an image space generator $G_{\mathrm{IM}}$ were cascaded in KIGAN. Adjacent undersampled \textit{k}-space slices (along the third dimension or the temporal dimension) $y^{l-1}_u$, $y^{l}_u$ and $y^{l+1}_u$ were the input of $G_{\mathrm{K}}$. A data consistency step, i.e., $\widetilde{y}_u=\bar{\Psi} G_{\mathrm{K}}(y^{l-1}_u,y^{l}_u,y^{l+1}_u) + y^{l}_u$ was applied for merging the output with the reconstructed missing data. $G_{\mathrm{IM}}$ was able to generate the output of $G_{\mathrm{K}}$ in image space, i.e., $\widetilde{x}_u$, to reconstructed image $\hat{x}_u$. Refinement learning $\hat{x}_u=G_{\mathrm{IM}}(\widetilde{x}_u)+\widetilde{x}_u$ was also used as the output of $G_{\mathrm{IM}}$ instead of $\hat{x}_u=G_{\mathrm{IM}}(\widetilde{x}_u)$. The reconstruction MR image $\hat{x}_u$, together with MR ground truth image $x_t$ were sent to the discriminator for the adversarial loss.

\begin{figure}[!h]
\centering\includegraphics[width=5in]{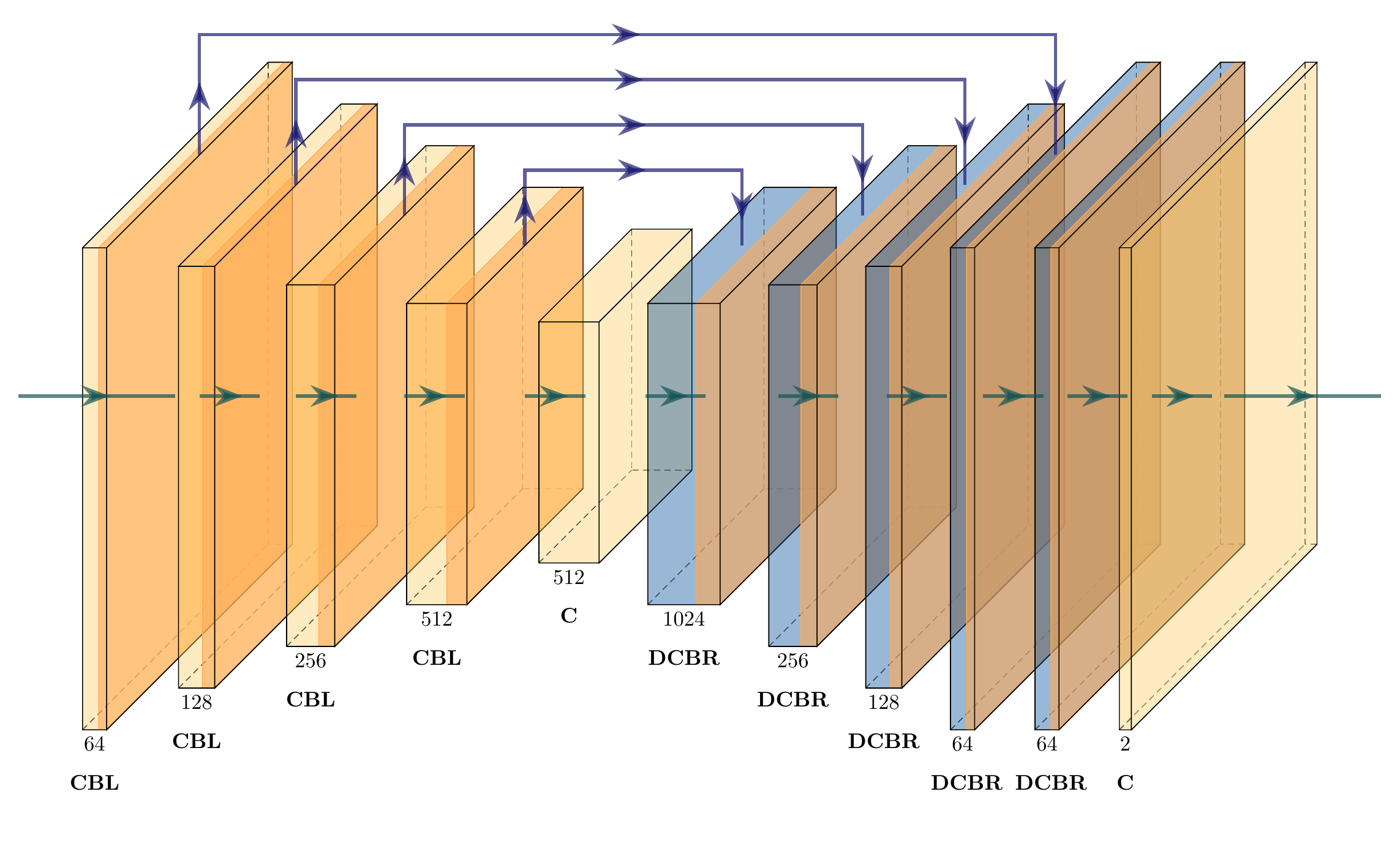}
\caption{The Structure of \textit{k}-space Generator in KIGAN. Five convolutional layers and five de-convolutional layers were applied as encoding path and decoding path in the generator respectively. A convolutional layer was used as the output layer to adjust the channel of the reconstructed MR image. Corresponding layers of the same scales were linked by the skip connection. (C: Convolutional layer; DC: De-Convolutional layer; B: Batch Normalization layer; L: Leaky ReLU layer; R: ReLU layer.)
}
\label{kigan_generator_K}
\end{figure}

\begin{figure}[!h]
\centering\includegraphics[width=5in]{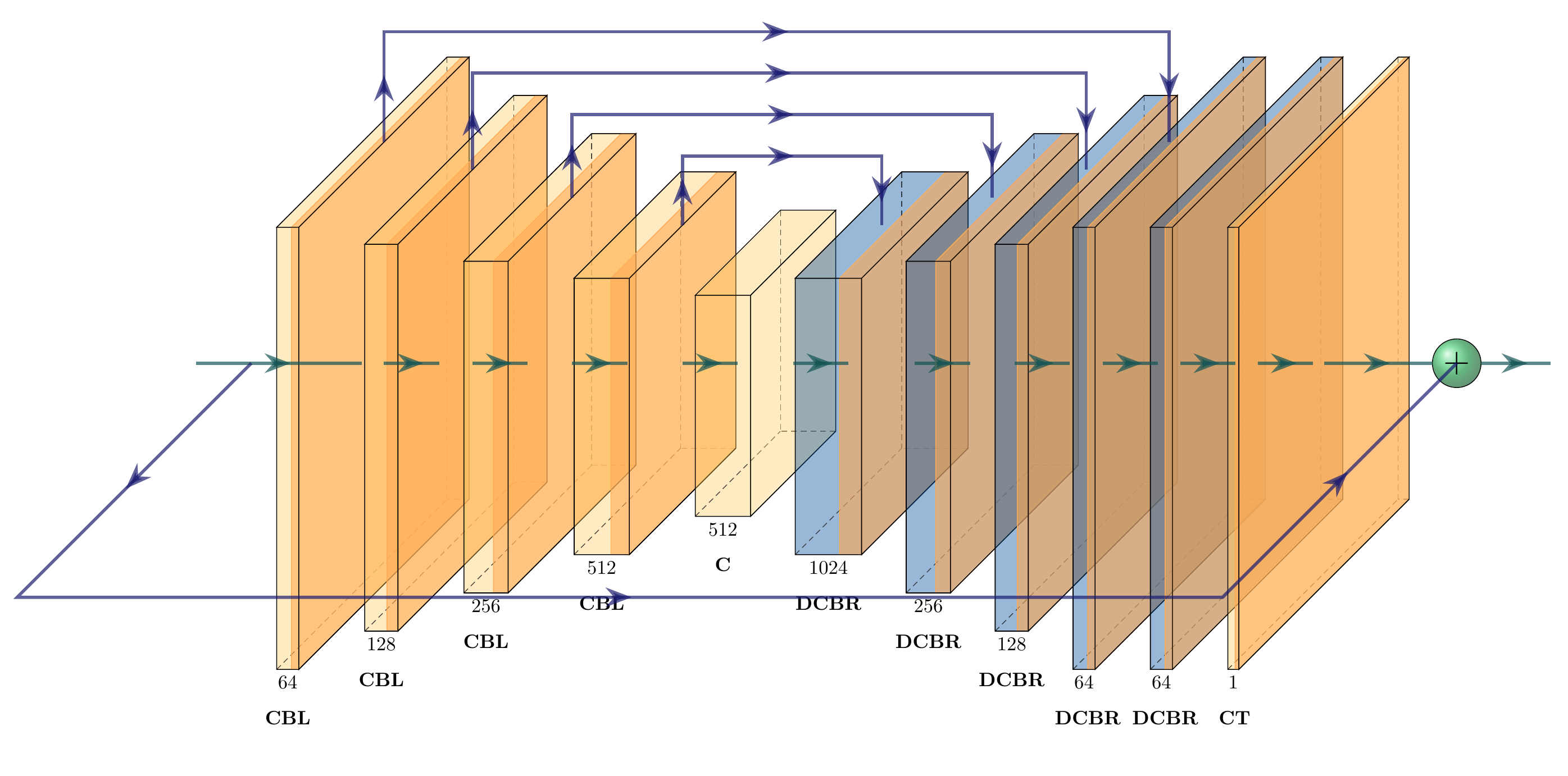}
\caption{The Structure of image space Generator in KIGAN. 5 convolutional layers and 5 de-convolutional layers were applied as encoding path and decoding path in the generator respectively. A convolutional layer and a hyperbolic tangent function were used as the output layer. Skip connection and shortcut connection were adopted in the image space generator. (C: Convolutional layer; DC: De-Convolutional layer; B: Batch Normalization layer; L: Leaky ReLU layer; R: ReLU layer; T: hyperbolic tangent function.)
}
\label{kigan_generator_i}
\end{figure}

As shown in Figures \ref{kigan_generator_K} and \ref{kigan_generator_i}, the main structure of the generators $G_{\mathrm{K}}$ and $G_{\mathrm{IM}}$ were based on a modified U-Net, which consisted of 5 convolutional layers in the encoding path and 5 deconvolutional layers in the decoding path, where each convolutional and deconvolutional layer were followed by a BN layer and a LReLU layer. Skip connection was applied between corresponding layers in the encoding and decoding paths, in order to pass the feature of different scales to the decoding layer for better reconstruction details. Additionally, in $G_{\mathrm{IM}}$, a shortcut connection between the input and output was applied for turning $G_{\mathrm{IM}}$ into a refinement function. 

\begin{figure}[!h]
\centering\includegraphics[width=5in]{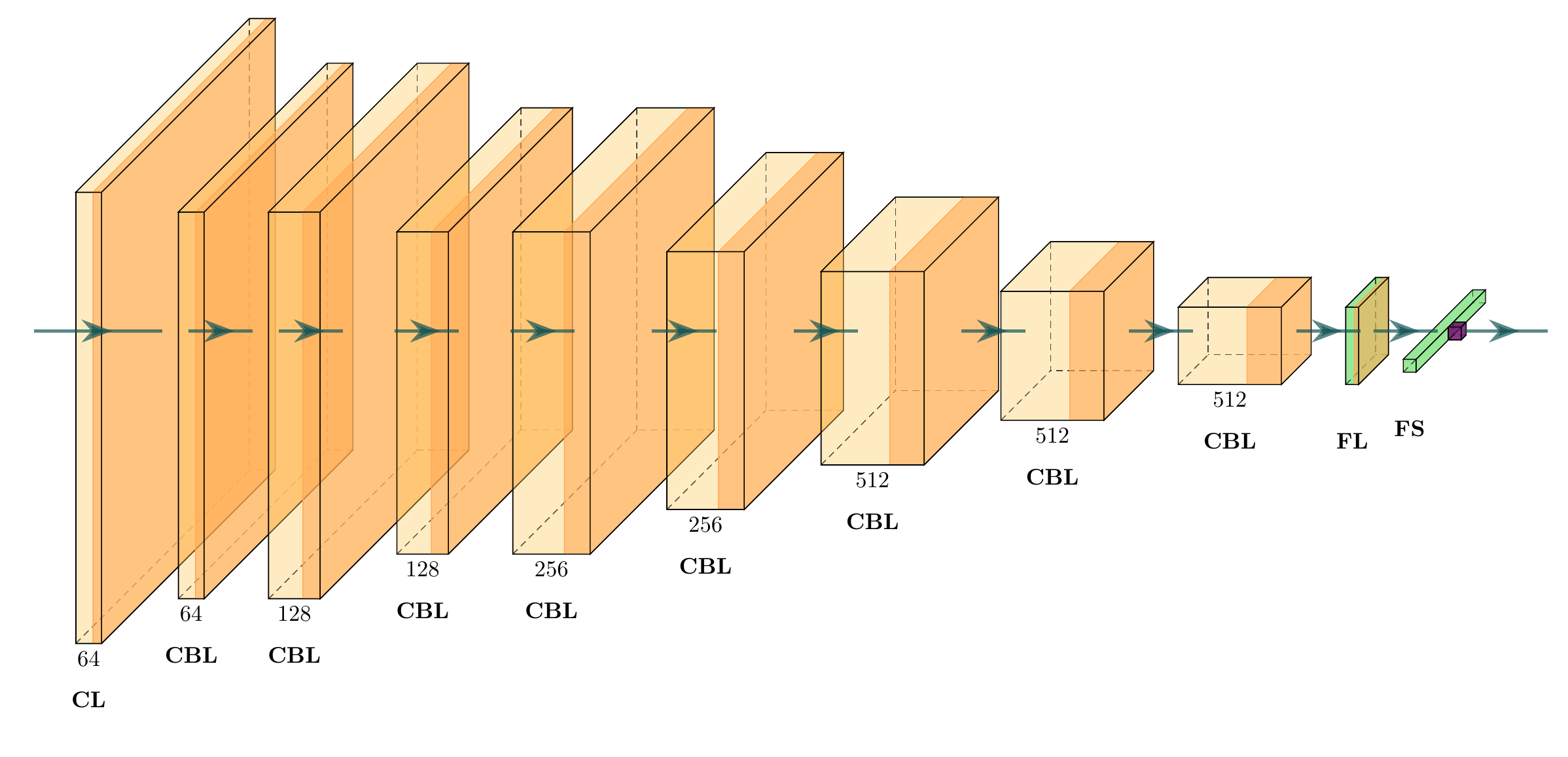}
\caption{The Structure of Discriminator in KIGAN. The discriminator was a standard 9-layer CNN structure with an FC layer and a Sigmoid activation function connected for the result of the classification. (C: Convolutional layer; B: Batch Normalization layer; L: Leaky ReLU layer; F: Full Connection layer; S: Sigmoid function.)}
\label{kigan_discriminator}
\end{figure}

As shown in Figure \ref{kigan_discriminator}, the discriminator was a standard 9-layer CNN structure with an FC layer and a Sigmoid activation function connected for the result of the classification. 

The loss function of KIGAN consisted of: an image domain MSE loss $L_{\mathrm{iMSE}}$, a frequency domain MSE loss $L_{\mathrm{fMSE}}$, and an adversarial loss $L_{\mathrm{adv}}$, which can be defined as:
\begin{equation}
L_{\mathrm{iMSE}}(\theta_{\mathrm{K}}, \theta_{\mathrm{IM}}) 
=\frac{1}{2}\mid\mid x_t -  \hat x_u \mid\mid^2_2,
\end{equation}

\begin{equation}
L_{\mathrm{fMSE}}(\theta_{\mathrm{K}}) 
=\frac{1}{2}\mid\mid y^l_t- \widetilde y_u \mid\mid^2_2,
\end{equation}

\begin{equation}
L_{\mathrm{adv}}(\theta_{\mathrm{K}}, \theta_{\mathrm{IM}}, \theta_D)
=\mathbb{E}_{x_t \sim M}
[\mathop{\text{log}} D_{\theta_D}(x_t)]
+\mathbb{E}_{\hat x_u \sim S}
[\mathop{\text{log}} (1-D_{\theta_D}(\hat x_u)],
\end{equation}
where $M$ denotes the collection of image space ground truth, and $S$ denotes the collection of image space reconstruction.

The whole loss function can be represented as:
\begin{dmath}
L_{\mathrm{TOTAL}}(\theta_{\mathrm{K}}, \theta_{\mathrm{IM}}, \theta_D) 
= \alpha L_{\mathrm{iMSE}}(\theta_{\mathrm{K}}, \theta_{\mathrm{IM}}) 
+ \beta L_{\mathrm{fMSE}}(\theta_{\mathrm{K}}) ~\\
+ L_{\mathrm{adv}}(\theta_{\mathrm{K}}, \theta_{\mathrm{IM}}, \theta_D), 
\end{dmath}
where $\alpha$, $\beta$ are the hyperparameters that control the balance of different components in the loss function.

\subsubsection{ReconGAN/RefineGAN}

\begin{figure}[!h]
\centering\includegraphics[width=5in]{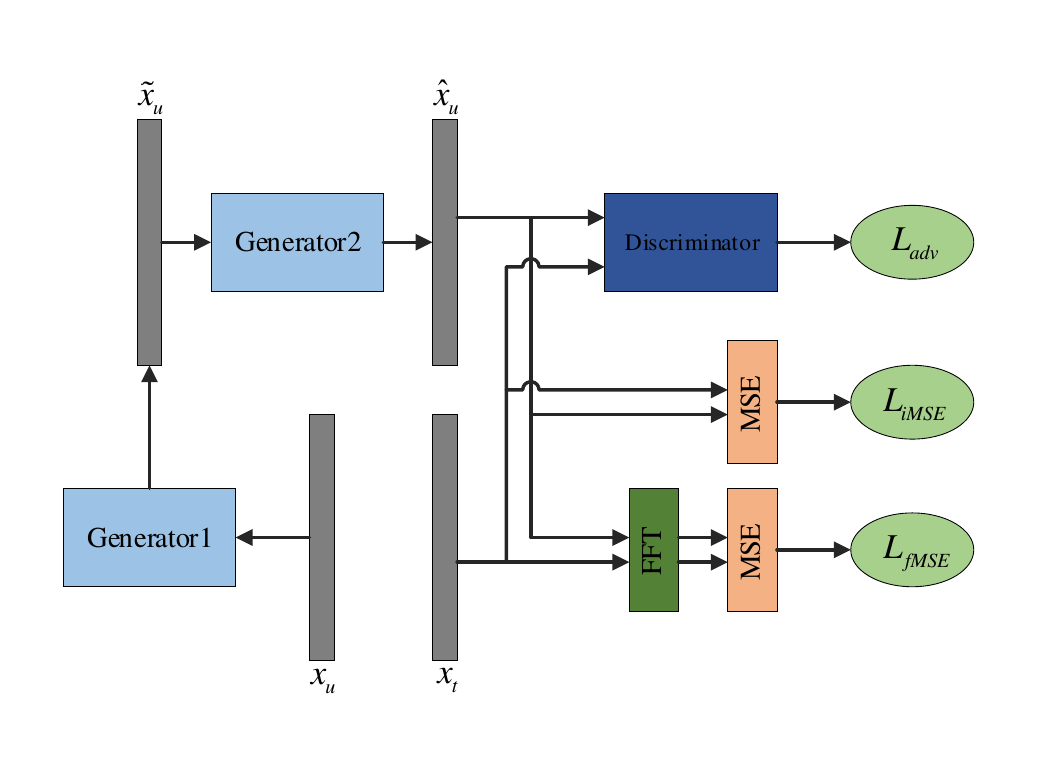}
\caption{The architecture of ReconGAN/RefineGAN. A two-fold chained network with 2 generators cascaded was used as the generator to produce reconstructed MR images $\hat x_u$ from undersampled MR images $x_u$. Reconstructed MR images $\hat x_u$ together with ground truth MR images $x_t$ were sent to the discriminator for the calculation of the adversarial loss $L_{\mathrm{adv}}$. Besides, image domain MSE loss $L_{\mathrm{iMSE}}$ and frequency domain MSE loss $L_{\mathrm{fMSE}}$ were added to the total loss function for training the generator.}
\label{reconrefinegan}
\end{figure}

ReconGAN/RefineGAN was proposed by Quan et al. \cite{Quan2018} for compressed sensing MRI reconstruction. Figure \ref{reconrefinegan} shows the architecture of ReconGAN/RefineGAN. The two-fold chained network that consisted of 2 U-shaped generators ($G_1$ and $G_2$ respectively) was able to generate a 2-channel zero-filling MR image (real part and imaginary part) directly into a 2-channel reconstructed MR image. The checkpoints after $G_1$ and $G_2$ were defined as ReconGAN and RefineGAN respectively.

\begin{figure}[!h]
\centering\includegraphics[width=5in]{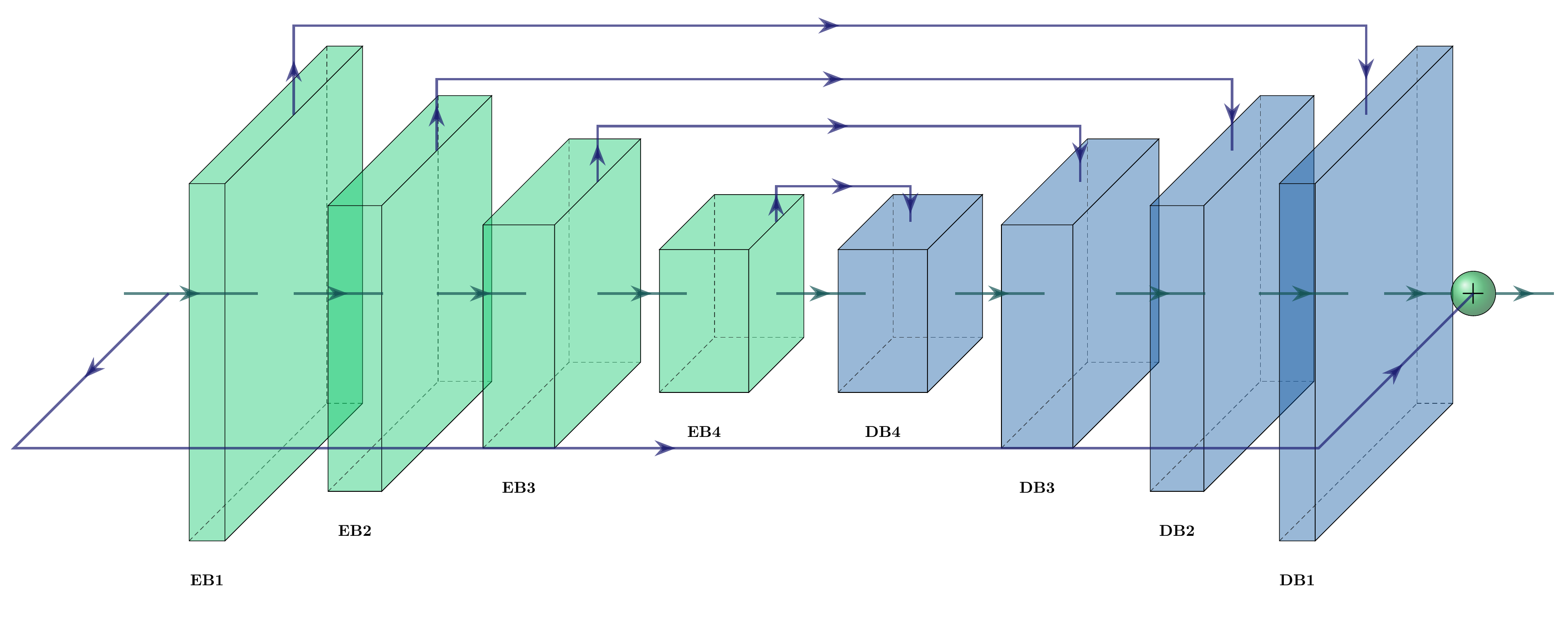}
\caption{Structure of the generator in ReconGAN/RefineGAN. The generator consisted of 4 encoder blocks in the encoding path and 4 decoder blocks in the decoding path. Skip connection was linked between blocks of the same scale in different paths. A shortcut connection was applied between the input and output of the generator, which turns the generator into a refinement function. (EB: Encoder Block; DB: Decoder Block.)
}
\label{reconrefinegan_generator}
\end{figure}

\begin{figure}[!h]
\centering\includegraphics[width=3in]{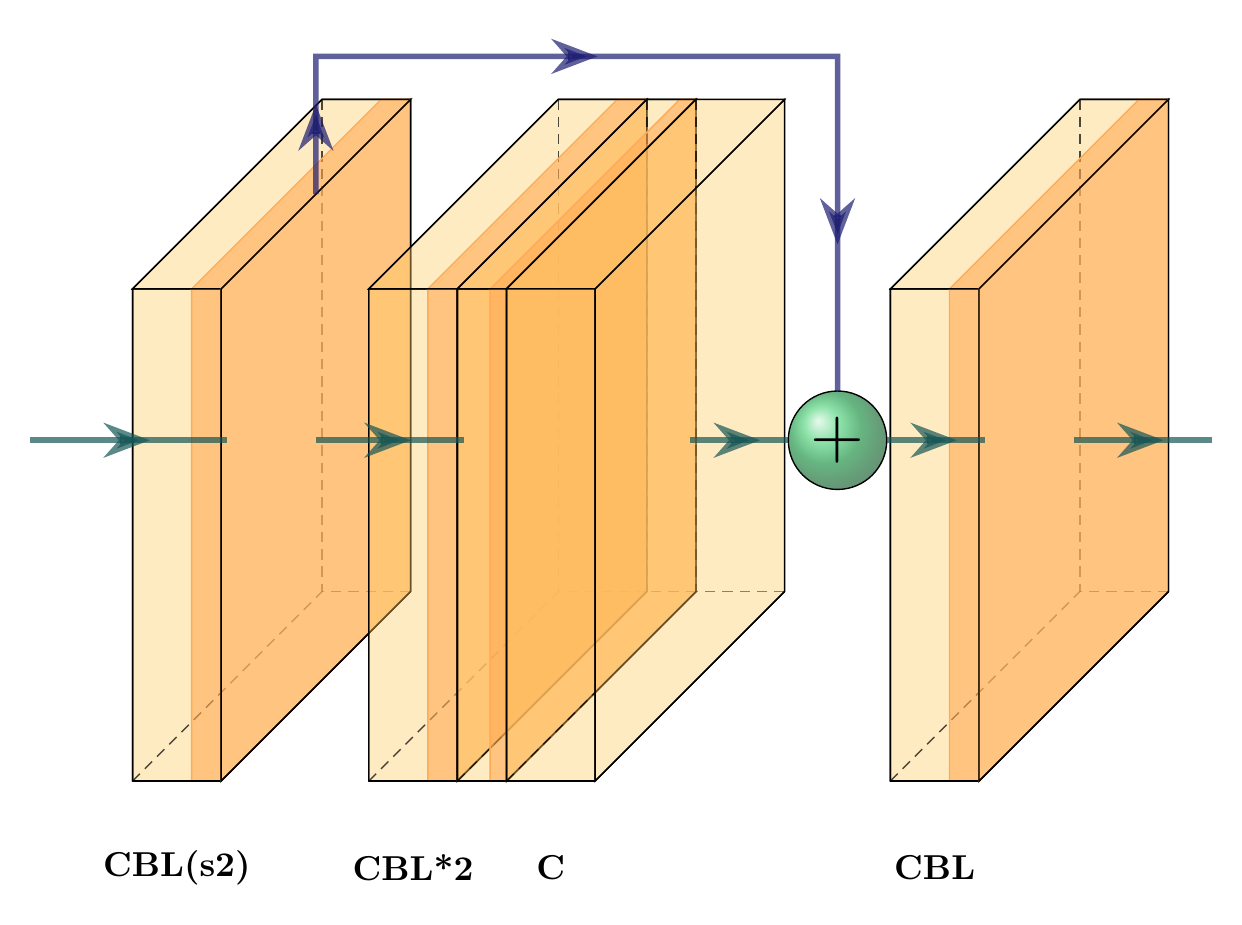}
\caption{Structure of the encoder block in the generator of ReconGAN/RefineGAN. (C: Convolutional layer; B: Batch Normalization layer; L: Leaky ReLU layer.)}
\label{reconrefinegan_encoder}
\end{figure}

\begin{figure}[!h]
\centering\includegraphics[width=3in]{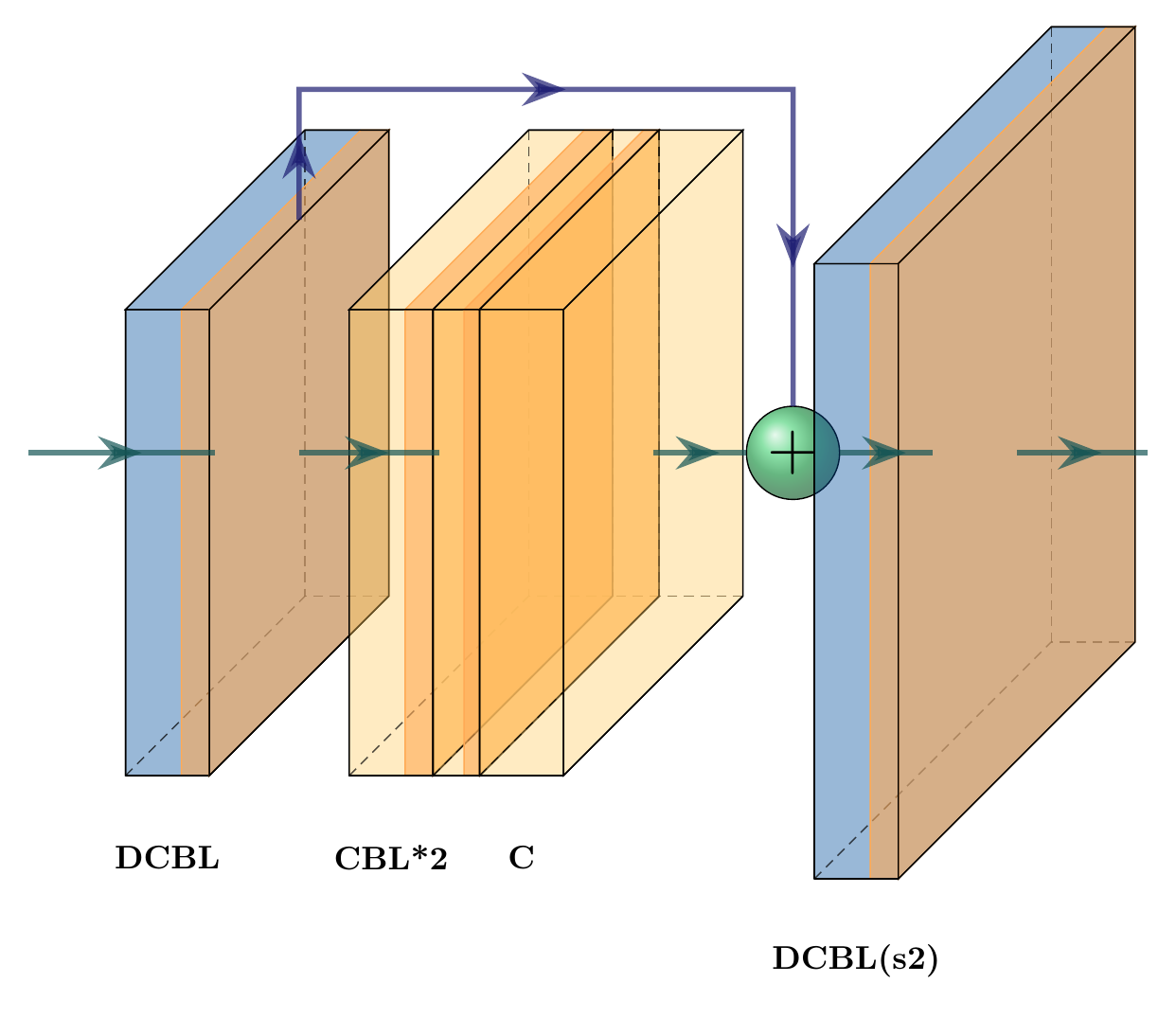}
\caption{Structure of the decoder block in the generator of ReconGAN/RefineGAN. (C: Convolutional layer; DC: De-Convolutional layer; B: Batch Normalization layer; L: Leaky ReLU layer.)}
\label{reconrefinegan_decoder}
\end{figure}

As shown in Figure \ref{reconrefinegan_generator}, the generator consisted of 4 encoder blocks in the encoding path and 4 decoder blocks in the decoding path. Skip connection between corresponding blocks that had the same scale was adopted to pass the feature from the encoding path to the decoding path. $\bar x_u=G_1(x_u)+x_u$ and $\hat x_u=G_2(\bar x_u)+\bar x_u$ were used as the output of the generator instead of $\bar x_u=G_1(x_u)$ and $\hat x_u=G_2(\bar x_u)$ for better reconstruction details and faster convergence speed. Figures \ref{reconrefinegan_encoder} and \ref{reconrefinegan_decoder} show the structure of the encoder block and decoder block. The encoder block consisted of 2 convolutional layers and a residual block inserted between them. The stride of the first convolutional layer was set to 2 for the downsampling. The decoder block consisted of 2 transposed convolutional layers and a residual block inserted between them. The stride of the second transposed convolutional layer was set to 2 for upsampling. 
The residual block consisted of three convolutional layers, the kernel number in the second convolutional layer was half of the kernel number in the first and third convolutional layer. A shortcut connection was linked between the input and output of the residual block.

\begin{figure}[!h]
\centering\includegraphics[width=3in]{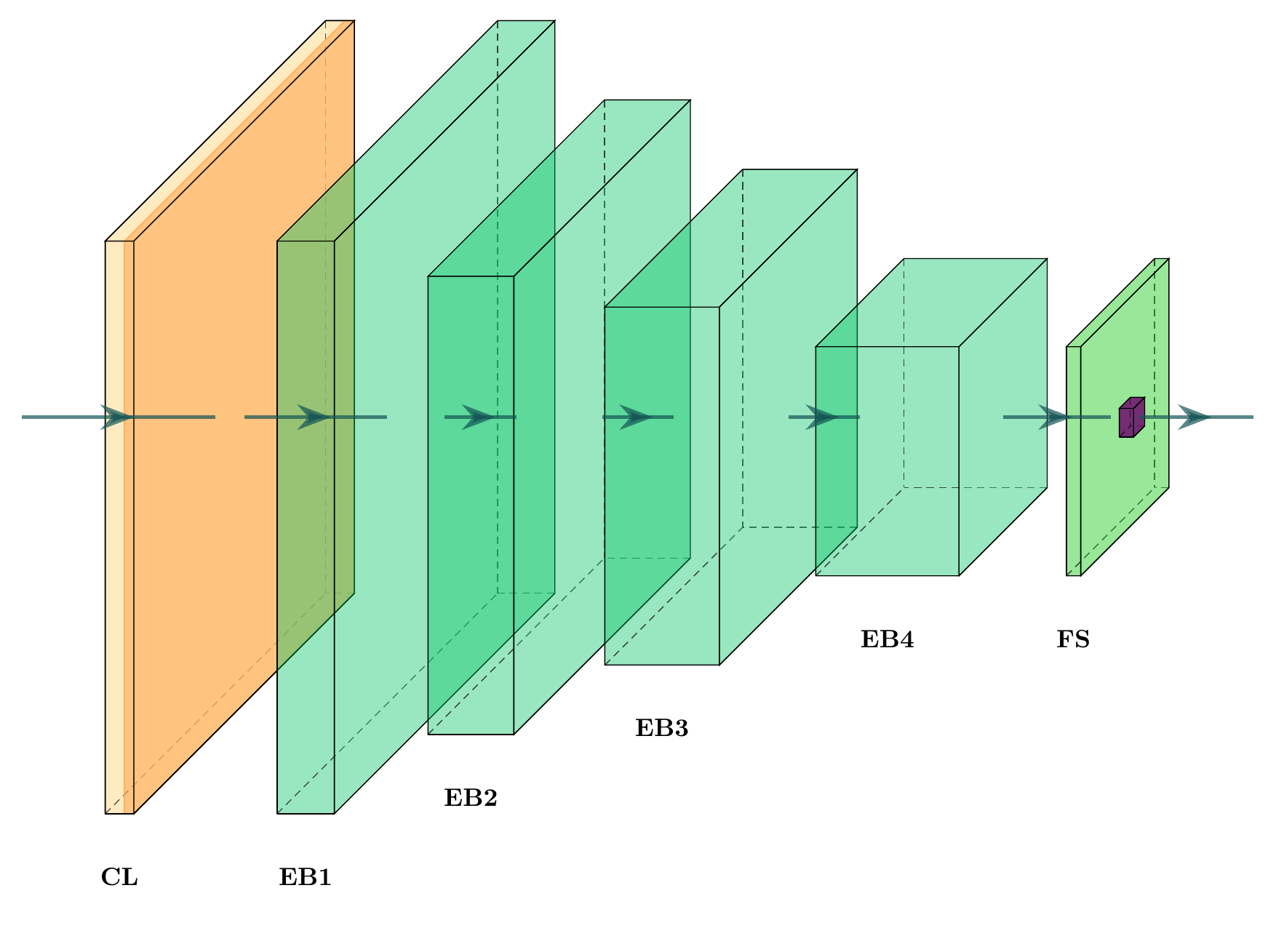}
\caption{Structure of the discriminator in ReconGAN/RefineGAN. The main structure adopted the encoding path of the generator. A full connection layer and a Sigmoid activation function were cascaded at the end of the discriminator. (C: Convolutional layer; L: Leaky ReLU layer; F: Full Connection layer; S: Sigmoid function; EB: Encoder Block.)}
\label{reconrefinegan_discriminator}
\end{figure}

As shown in Figure \ref{reconrefinegan_discriminator}, the encoding path in the generator was adopted as the main structure of the discriminator. An FC layer and Sigmoid function were cascaded after the encoding path for the classification result.

The loss function of ReconGAN/RefineGAN was consisted of: image domain MSE loss $L_{\mathrm{iMSE}}$, frequency domain MSE loss $L_{\mathrm{fMSE}}$, and the adversarial loss $L_{\mathrm{adv}}$, which can be defined as:

\begin{equation}
L_{\mathrm{iMSE}}(\theta_{G_1}, \theta_{G_2}) =
\frac{1}{2}\mid\mid x_t -  \hat x_u \mid\mid^2_2,
\end{equation}

\begin{equation}
L_{\mathrm{fMSE}}(\theta_{G_1}, \theta_{G_2}) =
\frac{1}{2}\mid\mid \mathcal{F} x_t- \mathcal{F} \hat x_u \mid\mid^2_2,
\end{equation}

\begin{dmath}
L_{\mathrm{adv}}(\theta_{G_1}, \theta_{G_2}, \theta_D)
=\mathbb{E}_{{x_t} \sim P_{train}(x_t)}
[\mathop{\text{log}} D_{\theta_D}({x_t})]
+\mathbb{E}_{{x_u} \sim p_{G_2G_1}(x_u)}
[\mathop{\text{log}} (1-D_{\theta_D}(G_{\theta_{G_2}}(G_{\theta_{G_1}}(x_u)))].
\end{dmath}

The total loss function can be represented as:

\begin{dmath}
L_{\mathrm{TOTAL}}(\theta_{G_1}, \theta_{G_2}, \theta_D)
= \alpha L_{\mathrm{iMSE}}(\theta_{G_1}, \theta_{G_2})
+ \beta L_{\mathrm{fMSE}}(\theta_{G_1}, \theta_{G_2})
+ L_{\mathrm{adv}}(\theta_{G_1}, \theta_{G_2}, \theta_D),
\end{dmath}
where $\alpha=10$, $\beta=0.1$ were the hyperparameters that controlled the balance of different components in the loss function. 

\subsection{Evaluation Methods}
Generally, the evaluation methods include objective methods for fidelity quality assessment and subjective methods for perceptual quality assessment. In this section, we review the most popular metrics for fast MRI quality evaluation.

\subsubsection{Fidelity Quality Assessment}
First, we introduce the Peak Signal-to-Noise Ratio (PSNR), which is the most commonly used evaluation criteria for image transformation tasks (e.g., reconstruction, super-resolution, de-noising). It involves the data range to measure the pixel-level Mean Squared Error (MSE):
\begin{equation}\label{eqt:psnr}
    \mathrm{PSNR}(I_{\mathrm{rec}}, I_{\mathrm{gt}}) = 10 \cdot \log_{10}(\frac{L^2}{\frac{1}{N}\sum_{i=1}^{N}(I_{\mathrm{rec}}(i)-I_{\mathrm{gt}}(i))^2}),
\end{equation}
where $L$ denotes the data range (generally $L = 1.0$ in MRI reconstruction tasks), and $N$ is the number of all the pixels in $I_{\mathrm{rec}}$ and $I_{\mathrm{gt}}$. PSNR represents the pixel-wise accuracy of the reconstruction regardless of the acquisition sequences of the multimodal MRI.

Besides, considering the importance of image structural information, such as brightness, contrast and structures, Structural SIMilarity index (SSIM) is formed as:
\begin{equation}\label{eqt:ssim}
    \mathrm{SSIM}(x, y) = \frac{2\mu_{x}\mu_{y} + \kappa_1}{\mu_x^2 + \mu_y^2 + \kappa_1} \cdot \frac{\sigma_{xy}+\kappa_2}{\sigma_x^2 + \sigma_y^2 + \kappa_2}, 
\end{equation}
where $x, y$ denote two images, $\mu$ and $\sigma^2$ are the mean and variance, $\sigma_{xy}$ is the covariance between $x$ and $y$, and $\kappa_1, \kappa_2$ are constant relaxation terms.

\subsubsection{Perceptual Quality Assessment}
The perceptual quality of an image represents how realistic it looks. In MRI images reconstruction tasks, the most reliable perceptual quality assessment is the mean opinion score (MOS), which asks experienced radiologists to rate the reconstructed images. Typically, the images are rated from 0 to 4 depending on the reconstructed image quality (i.e., non-diagnostic, poor, fair, good, and excellent), and the final MOS is calculated as the arithmetic mean of the scores of all raters. In some cases, the rater may also mark the low perceptual quality features such as low SNR and motion artefacts. Although the MOS seems to be faithful, it has limitations such as inter-/inner-raters bias and variance of rating criteria and the scoring might be time-consuming. Thus, the Frechet Inception Distance (FID) \cite{FID}, as a learning based perceptual quality assessment, is becoming more commonly used for evaluation in GAN based image reconstruction tasks. It considers the high-level global features of a group of images (e.g., the reconstructed images) as a multidimensional Gaussian distribution $\mathcal{N}(\mu, \Sigma)$, and measures the differences between the two distributions of the reconstructed images $\mathbb{I}_{\mathrm{rec}}$ and the ground truth images $\mathbb{I}_{\mathrm{gt}}$. It first converts each group of images into a distribution of 2048 features in the latent space of a pre-trained image classification model Inception-V3 \cite{InceptionV3}. Then, the FID between these two distributions is calculated as:
\begin{equation}\label{eqt:fid}
    \mathrm{FID}(\mathbb{I}_{\mathrm{rec}}, \mathbb{I}_{\mathrm{gt}}) = \left \|\mu_{\mathrm{gt}} - \mu_{\mathrm{rec}}\right \|^2 + \mathrm{Tr}(\Sigma_{\mathrm{gt}} +\Sigma_{\mathrm{rec}} -2(\Sigma_{\mathrm{gt}}\Sigma_{\mathrm{rec}})^{1/2}).
\end{equation}
The FID becomes a popular metric for image perceptual quality assessment in GAN based image generation tasks because it is fully automatic and the features extracted from Inception-V3 are close to real-world object classification problems, which tend to mimic human perception similarity for images.


\section{Benchmarking}
\begin{figure}[!h]
\centering\includegraphics[width=5in]{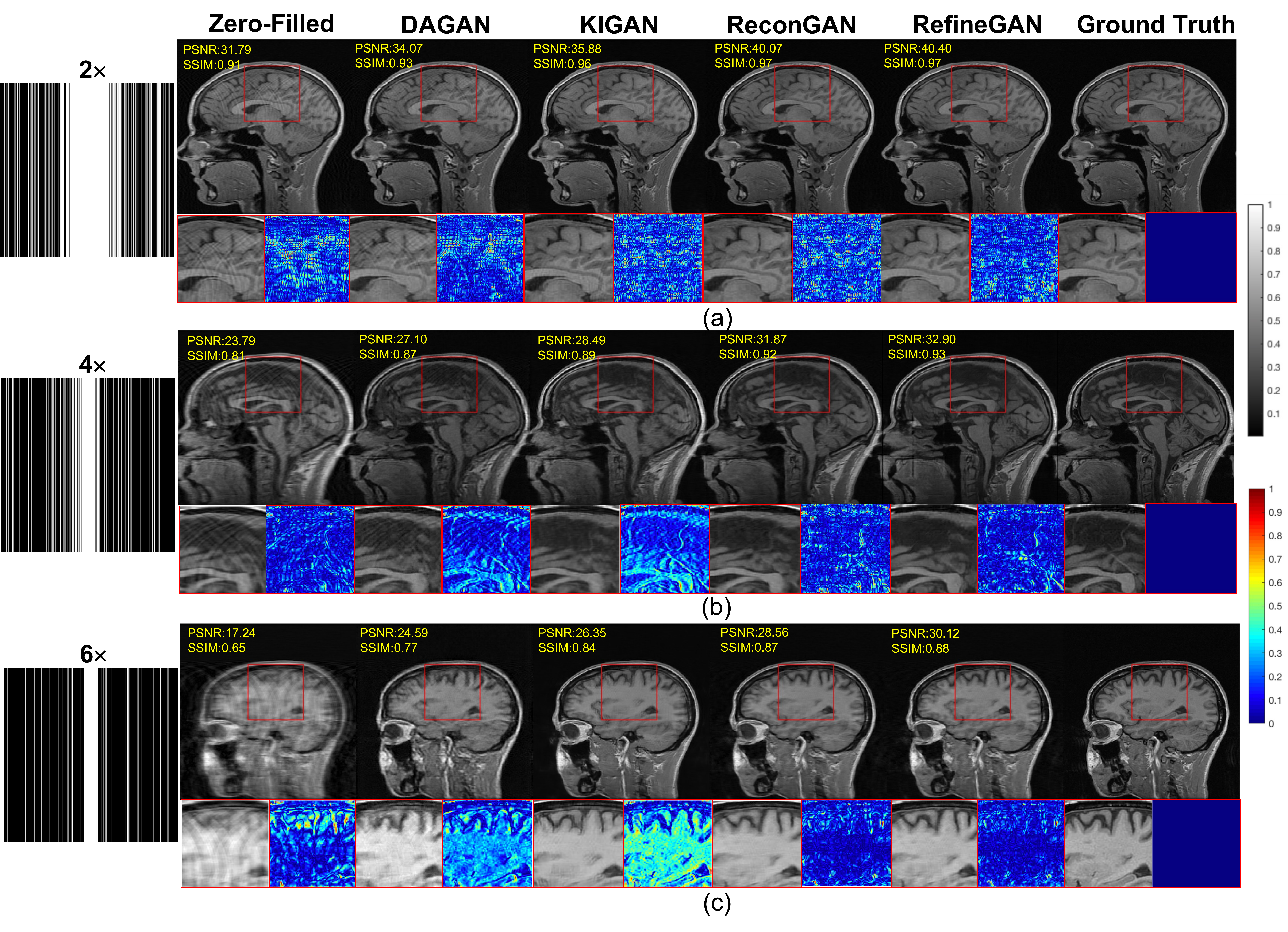}
\caption{Brain reconstruction result using the Cartesian mask. From top to bottom 2$\times$, 4$\times$ and 6$\times$ acceleration, respectively. From left to right are Zero-Filled (ZF), DAGAN, KIGAN, ReconGAN, RefineGAN, Ground Truth (GT).}
\label{brain_cartes}
\end{figure}

In this chapter, we benchmark four GAN based algorithms, i.e., DAGAN, KIGAN, ReconGAN and RefineGAN, for fast MRI. Figure \ref{brain_cartes} shows the brain reconstruction results using different acceleration factors (2$\times$, 4$\times$, 6$\times$). It is obvious that the zero-filled (ZF) image has strong artefacts inside the brain tissue. From the entire image, the DAGAN effectively removes the artefacts in the ZF. However, in terms of the zoomed-in areas, there still exists some residual artefacts. For the reconstructions produced by KIGAN, blurring artefacts still exist. Although ReconGAN shows a significant reduction of aliasing artefacts, the edge details are not reconstructed clearly enough. It can be seen that the reconstructed details of RefineGAN are relatively fine, and the reconstruction quality is close to that of the ground truth. 

\begin{figure}[!h]
\centering\includegraphics[width=5in]{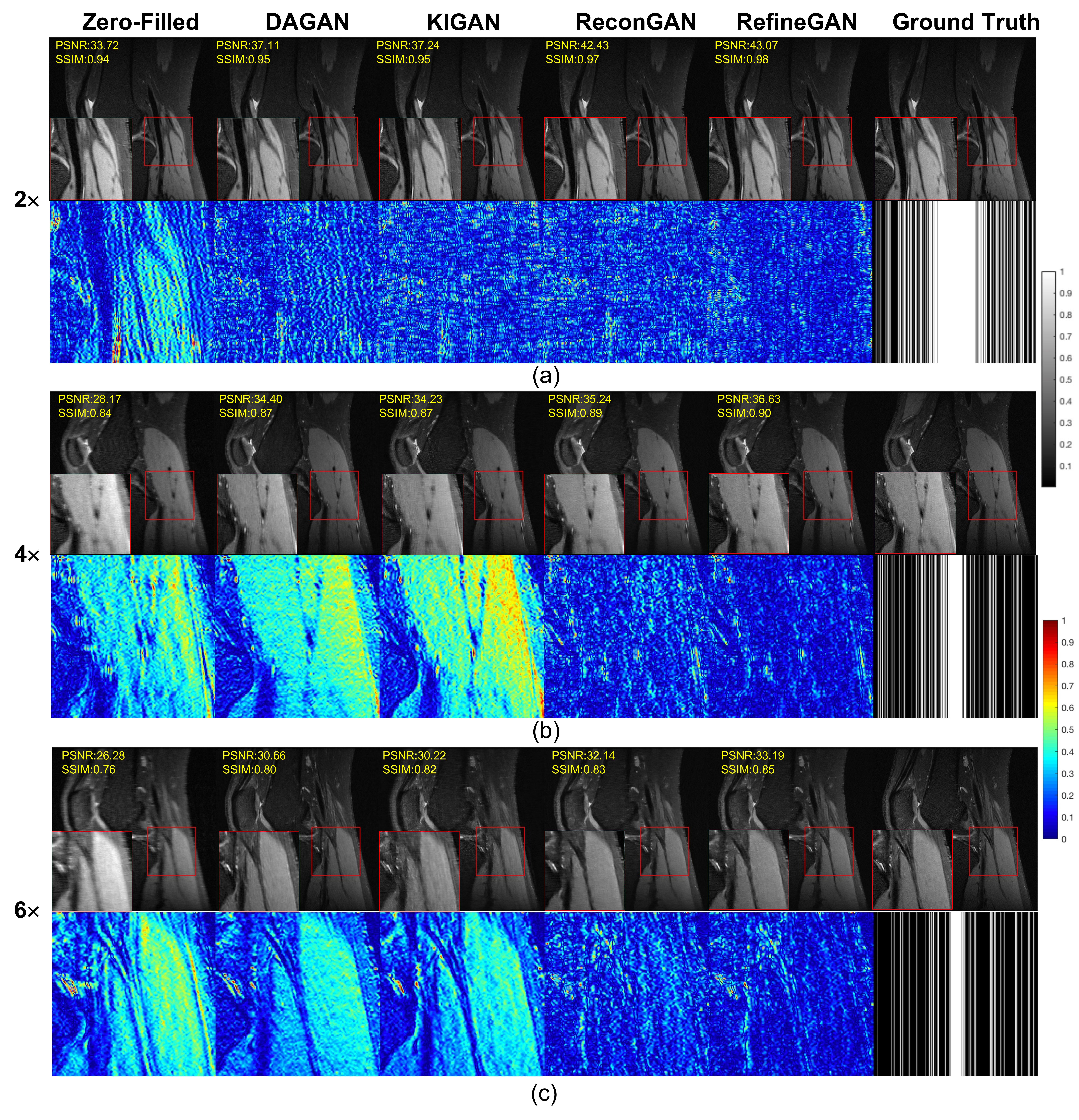}
\caption{Knee reconstruction result using the Cartesian mask. From top to bottom 2$\times$, 4$\times$ and 6$\times$ acceleration, respectively. From left to right are Zero-Filled (ZF), DAGAN, KIGAN, ReconGAN, RefineGAN, Ground Truth (GT).}
\label{knee_cartes}
\end{figure}

Besides, knee reconstruction results using different Cartesian masks are shown in Figure \ref{knee_cartes}. It can be seen that except for the ZF images, all methods can reconstruct acceptable MR images. As the acceleration factor goes high, obvious aliasing artefacts are produced in DAGAN images. As can be observed in the zoomed-in areas and the corresponding error maps, KIGAN can not restore clear vessels. ReconGAN and RefineGAN show better reconstruction results with higher PSNR and SSIM. In addition, the quantitative values of the RefineGAN are superior to those of the other methods.

\begin{figure}[!h]
\centering\includegraphics[width=5in]{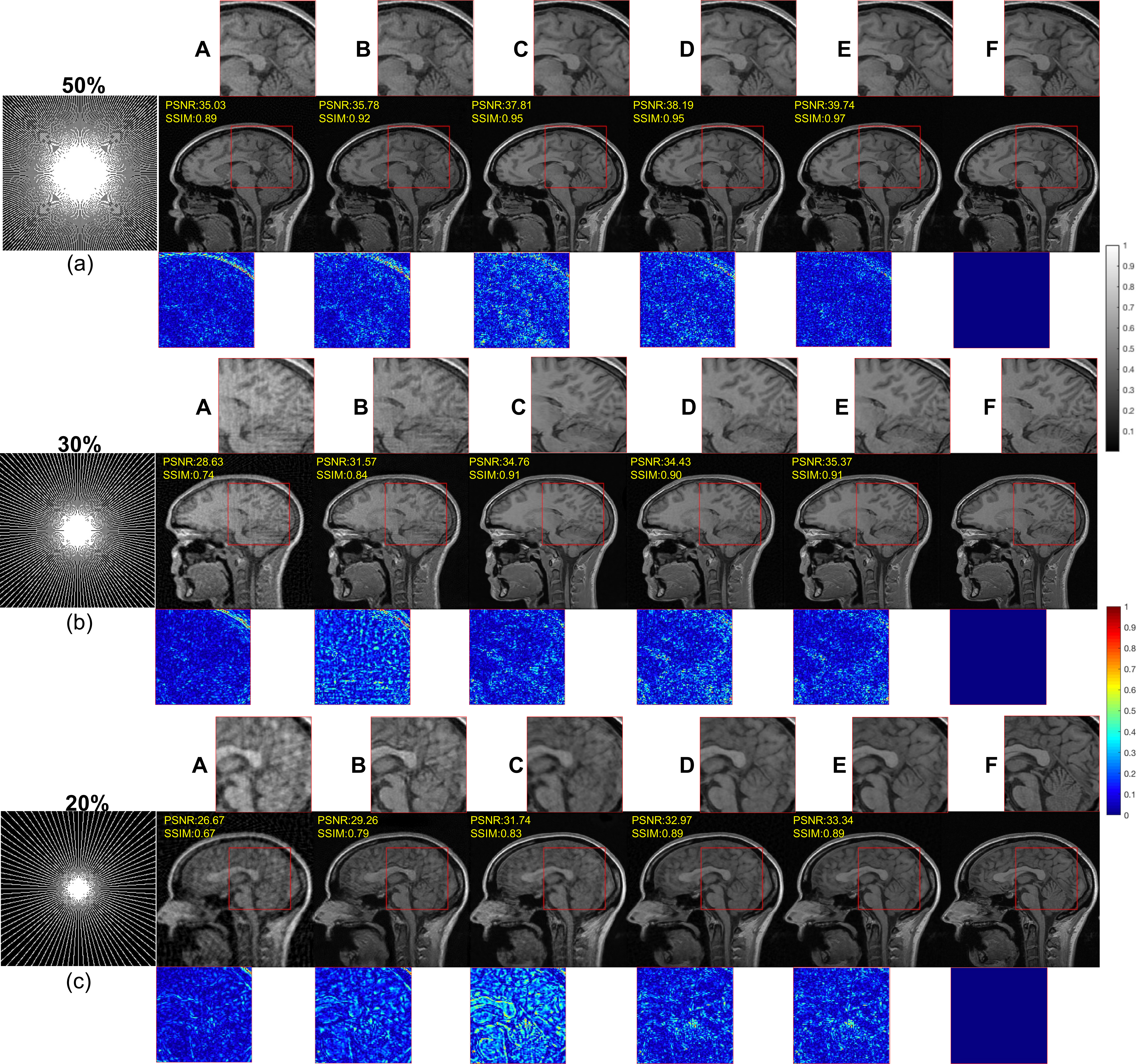}
\caption{Brain reconstruction result using the radial mask. From top to bottom, the sampling rate (SR) is 50\%, 30\%, and 20\%, respectively. From left to right are Zero-Filled (ZF), DAGAN, KIGAN, ReconGAN, RefineGAN, Ground Truth (GT).}
\label{brain_radial}
\end{figure}

\begin{figure}[!h]
\centering\includegraphics[width=5in]{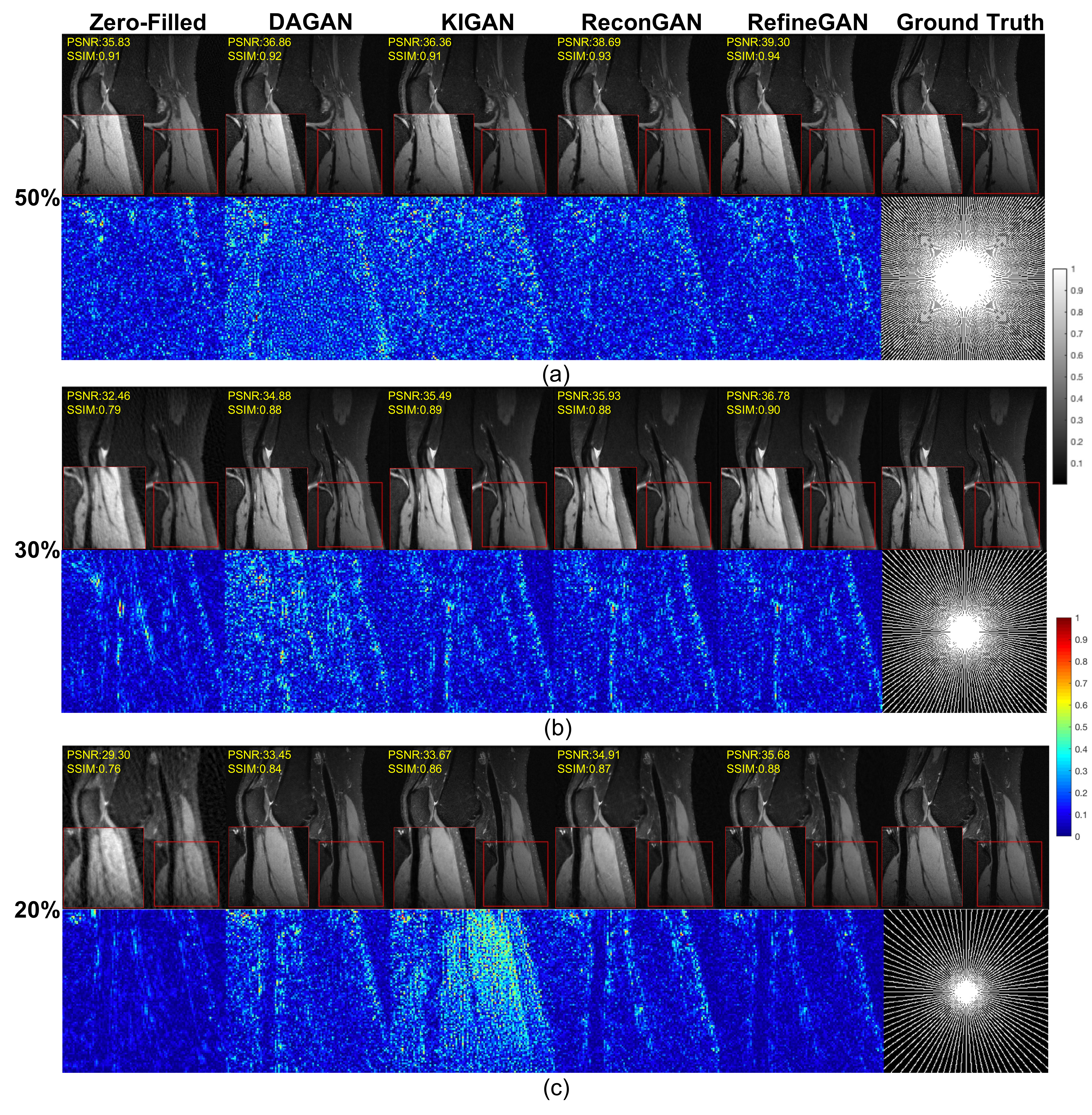}
\caption{Knee reconstruction result using the radial mask. From top to bottom 2$\times$, 4$\times$ and 6$\times$ acceleration, respectively. From left to right are Zero-Filled (ZF), DAGAN, KIGAN, ReconGAN, RefineGAN, Ground Truth (GT).}
\label{knee_radial}
\end{figure}

Furthermore, we also used radial and spiral masks for training and testing each GAN based method. The sampling rate (SR) of each mask is 50\%, 30\% and 20\%. Figures \ref{brain_radial} and \ref{knee_radial} show the brain and knee reconstruction results using radial masks. We can see that the image reconstructed by the ZF method under the radial mask has strong blurring artefacts, and the details in the brain and knee cannot be distinguished clearly. When SR=20\%, from the error maps, we can see that there are still obvious blurring artefacts and obscure blood vessels in the results of DAGAN and KIGAN. However, both ReconGAN and RefineGAN can restore sharper vessel edges and finer textures compared to other methods. Besides, RefineGAN has better PSNR and SSIM quantification.

\begin{figure}[!h]
\centering\includegraphics[width=5in]{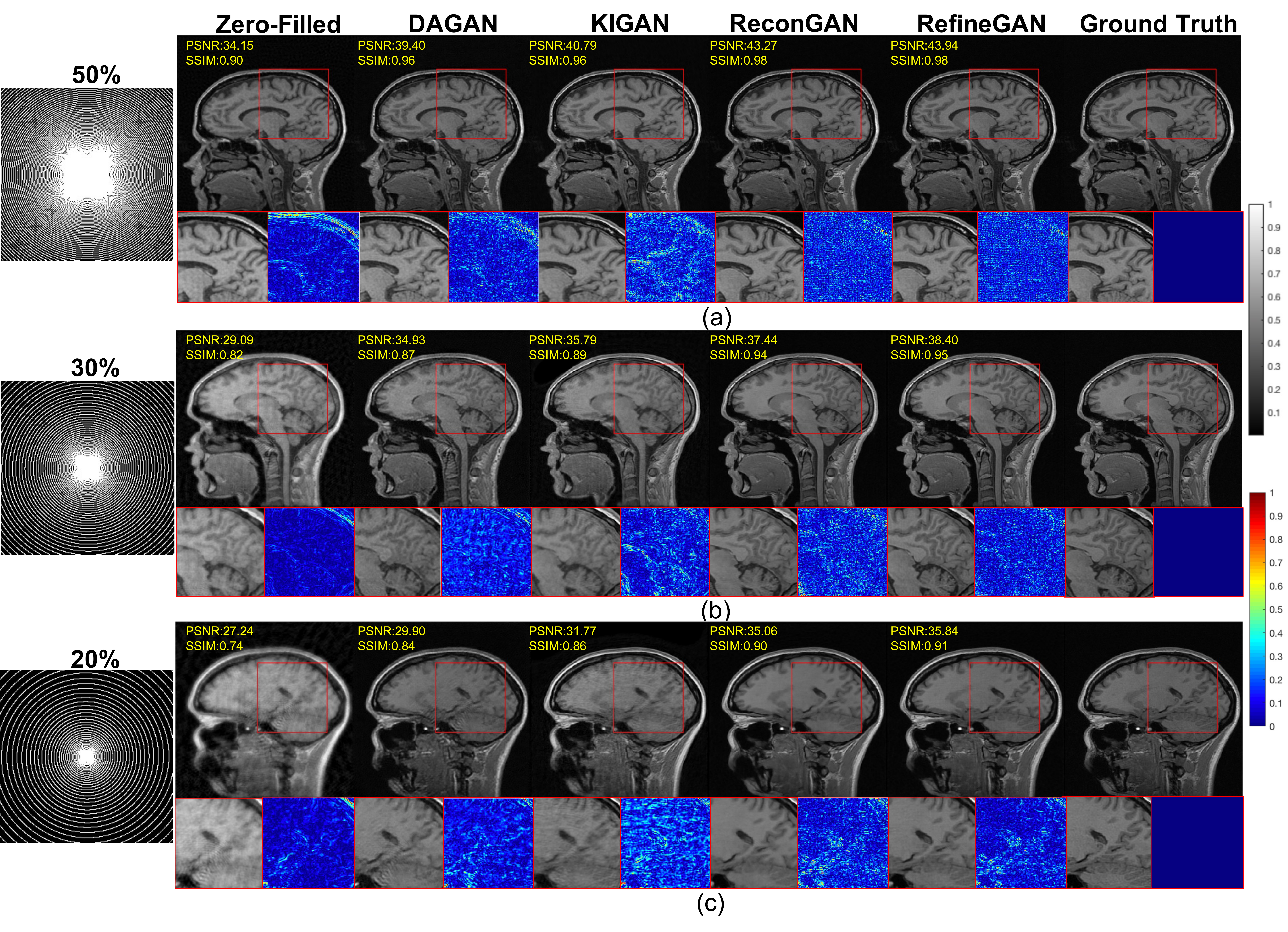}
\caption{Brain reconstruction result using the radial mask. From top to bottom, the sampling rate (SR) is 50\%, 30\%, and 20\%, respectively. From left to right are Zero-Filled (ZF), DAGAN, KIGAN, ReconGAN, RefineGAN, Ground Truth (GT).}
\label{brain_spiral}
\end{figure}

\begin{figure}[!h]
\centering\includegraphics[width=5in]{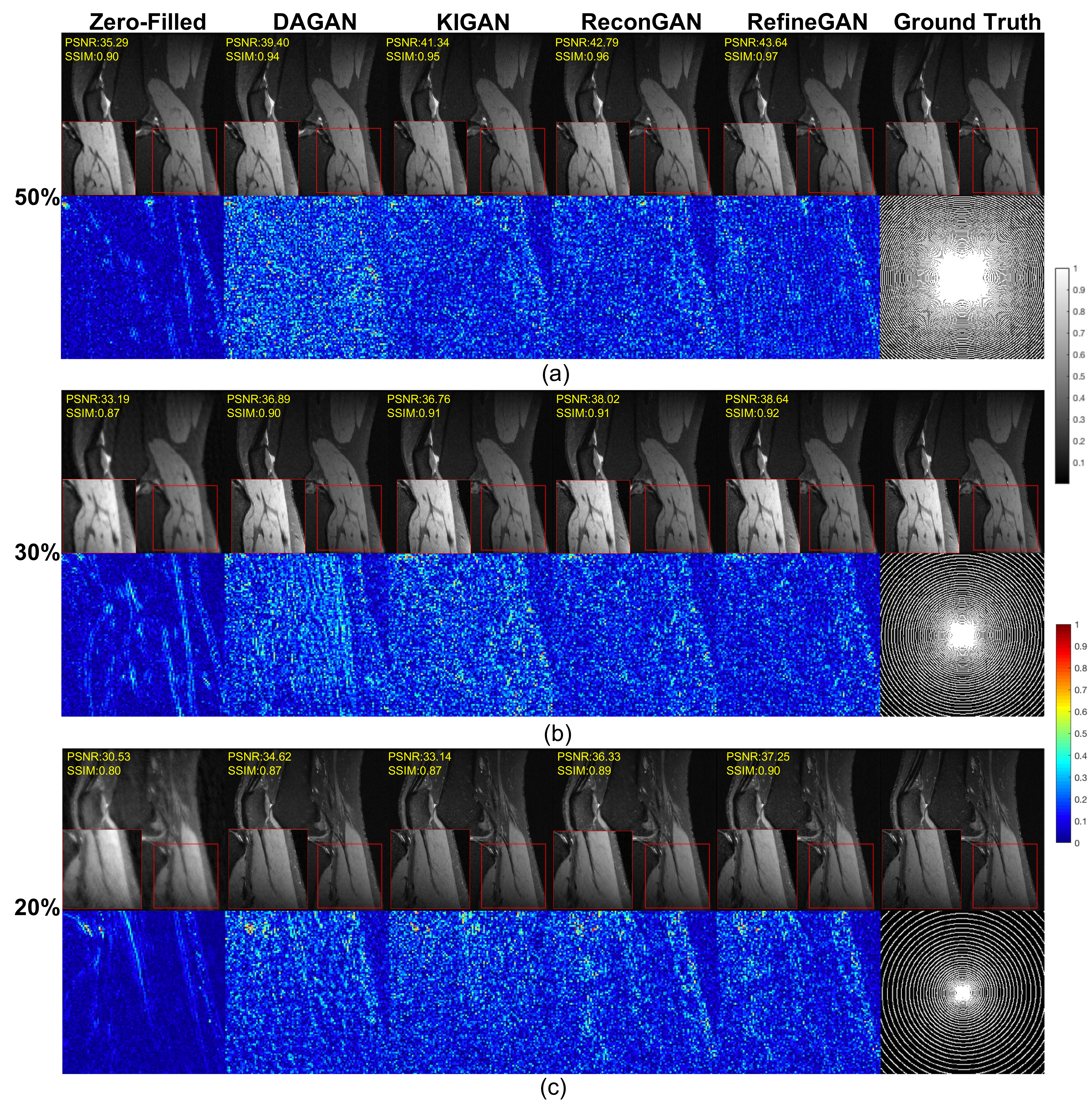}
\caption{Knee reconstruction result using the radial mask. From top to bottom 2$\times$, 4$\times$ and 6$\times$ acceleration, respectively. From left to right are Zero-Filled (ZF), DAGAN, KIGAN, ReconGAN, RefineGAN, Ground Truth (GT).}
\label{knee_spiral}
\end{figure}

The results are similar using spiral masks. Figures \ref{brain_spiral} and \ref{knee_spiral} show the brain and knee reconstruction results of each method using different spiral masks. Through the error map, we can intuitively see that the reconstructed image of RefineGAN has fewer errors, and the reconstruction details are also better. The image reconstructed by RefineGAN can clearly show the details of the gray matter in the brain and the details of the blood vessels in the knee. 

\begin{table}[]
\caption{The quantitative metrics (PSNR, SSIM, and RMSE($\times10^{-2}$)) of the brain using different GAN based methods). The bold numbers indicate the best results.}
\label{tab:my-table}
\resizebox{\textwidth}{!}{%
\begin{tabular}{cccccccc}
\hline
Mask                    & AF/SR                 & Metric & ZF         & DAGAN      & KIGAN      & ReconGAN           & \textbf{RefineGAN}  \\ \hline
\multirow{9}{*}{Cartesian} & \multirow{3}{*}{2X} & PSNR & 30.94±2.75 & 33.79±1.88 & 33.90±2.55 & 39.08±1.34 & \textbf{39.40±1.33} \\ \cline{3-8} 
                        &                       & SSIM   & 0.92±0.02  & 0.93±0.01  & 0.96±0.01  & \textbf{0.97±0.00} & \textbf{0.97±0.00}  \\ \cline{3-8} 
                        &                       & RMSE   & 1.57±0.01  & 0.72±0.43  & 0.78±0.53  & 0.20±0.09          & \textbf{0.19±0.08}  \\ \cline{2-8} 
                        & \multirow{3}{*}{4X}   & PSNR   & 23.69±3.02 & 28.76±1.95 & 28.14±1.84 & 32.07±1.65         & \textbf{32.67±1.56} \\ \cline{3-8} 
                        &                       & SSIM   & 0.79±0.03  & 0.86±0.02  & 0.88±0.02  & 0.92±0.01          & \textbf{0.93±0.01}  \\ \cline{3-8} 
                        &                       & RMSE   & 8.86±7.01  & 2.35±1.50  & 2.67±1.40  & 1.05±0.54          & \textbf{0.91±0.08}  \\ \cline{2-8} 
                        & \multirow{3}{*}{6X}   & PSNR   & 19.47±2.31 & 25.4±1.57  & 27.91±1.57 & 29.23±1.68         & \textbf{29.95±1.61} \\ \cline{3-8} 
                        &                       & SSIM   & 0.66±0.04  & 0.77±0.03  & 0.86±0.02  & 0.88±0.02          & \textbf{0.89±0.02}  \\ \cline{3-8} 
                        &                       & RMSE   & 21.3±14.1  & 4.89±2.33  & 2.75±1.31  & 2.03±1.00          & \textbf{1.71±0.83}  \\ \hline
\multirow{9}{*}{Radial} & \multirow{3}{*}{50\%} & PSNR   & 34.28±1.03 & 35.24±1.14 & 38.93±1.10 & 37.89±0.93         & \textbf{39.38±0.88} \\ \cline{3-8} 
                        &                       & SSIM   & 0.87±0.02  & 0.91±0.01  & 0.96±0.01  & 0.95±0.01          & \textbf{0.97±0.00}  \\ \cline{3-8} 
                        &                       & RMSE   & 1.94±0.22  & 1.74±0.22  & 1.01±0.12  & 1.28±0.13          & \textbf{0.88±0.11}  \\ \cline{2-8} 
                        & \multirow{3}{*}{30\%} & PSNR   & 28.83±0.72 & 30.99±2.33 & 33.97±1.05 & 34.35±0.99         & \textbf{35.21±1.05} \\ \cline{3-8} 
                        &                       & SSIM   & 0.73±0.02  & 0.82±0.02  & 0.91±0.01  & 0.91±0.01          & \textbf{0.92±0.01}  \\ \cline{3-8} 
                        &                       & RMSE   & 3.10±0.30  & 2.01±0.27  & 1.60±0.18  & 1.93±0.21          & \textbf{1.02±0.17}  \\ \cline{2-8} 
                        & \multirow{3}{*}{20\%} & PSNR   & 26.47±0.64 & 28.96±1.90 & 31.87±0.94 & 31.96±0.87         & \textbf{32.83±0.94} \\ \cline{3-8} 
                        &                       & SSIM   & 0.65±0.02  & 0.78±0.02  & 0.84±0.03  & 0.87±0.02          & \textbf{0.89±0.02}  \\ \cline{3-8} 
                        &                       & RMSE   & 6.08±0.36  & 2.47±0.45  & 2.56±0.27  & 2.54±0.25          & \textbf{1.56±0.32}  \\ \hline
\multirow{9}{*}{Spiral} & \multirow{3}{*}{50\%} & PSNR   & 34.87±1.01 & 38.08±0.99 & 41.56±1.00 & 43.69±0.58         & \textbf{44.36±0.57} \\ \cline{3-8} 
                        &                       & SSIM   & 0.90±0.01  & 0.95±0.01  & 0.96±0.01  & 0.97±0.01          & \textbf{0.98±0.00}  \\ \cline{3-8} 
                        &                       & RMSE   & 1.62±0.18  & 1.25±0.01  & 0.84±0.10  & 0.72±0.66          & \textbf{0.59±0.61}  \\ \cline{2-8} 
                        & \multirow{3}{*}{30\%} & PSNR   & 29.55±0.71 & 34.87±2.79 & 36.71±1.09 & 37.93±0.78         & \textbf{38.61±0.82} \\ \cline{3-8} 
                        &                       & SSIM   & 0.83±0.01  & 0.87±0.01  & 0.91±0.02  & 0.95±0.01          & \textbf{0.95±0.00}  \\ \cline{3-8} 
                        &                       & RMSE   & 3.93±0.28  & 1.72±0.41  & 1.65±0.20  & 1.27±0.11          & \textbf{0.94±0.19}  \\ \cline{2-8} 
                        & \multirow{3}{*}{20\%} & PSNR   & 26.62±0.63 & 28.87±2.29 & 32.43±0.72 & 35.02±0.81         & \textbf{35.11±0.85} \\ \cline{3-8} 
                        &                       & SSIM   & 0.73±0.01  & 0.80±0.04  & 0.88±0.03  & 0.91±0.01          & \textbf{0.92±0.01}  \\ \cline{3-8} 
                        &                       & RMSE   & 7.12±0.36  & 2.29±0.31  & 2.40±0.20  & 1.78±0.16          & \textbf{1.52±0.27}  \\ \hline
\end{tabular}%
}
\label{table_brain}
\end{table}

\begin{table}[]
\caption{The quantitative metrics (PSNR, SSIM, and RMSE($\times10^{-2}$)) of the knee using different GAN based methods. The bold numbers indicate the best results.}
\label{tab:my-table}
\resizebox{\textwidth}{!}{%
\begin{tabular}{cccccccc}
\hline
Mask                    & AF/SR                 & Metric & ZF          & DAGAN      & KIGAN      & ReconGAN           & \textbf{RefineGAN}  \\ \hline
\multirow{9}{*}{Cartesian} & \multirow{3}{*}{2X} & PSNR & \multicolumn{1}{l}{34.66±2.98} & \multicolumn{1}{l}{38.91±1.59} & 38.53±2.51 & 42.37±1.60 & \textbf{42.41±1.98} \\ \cline{3-8} 
                        &                       & SSIM   & 0.95±0.01   & 0.94±0.01  & 0.96±0.01  & 0.97±0.00          & \textbf{0.98±0.00}  \\ \cline{3-8} 
                        &                       & RMSE   & 1.64±1.32   & 0.52±0.31  & 0.83±1.03  & \textbf{0.23±0.09} & 0.24±0.16           \\ \cline{2-8} 
                        & \multirow{3}{*}{4X}   & PSNR   & 27.31±3.23  & 34.35±1.77 & 34.70±1.74 & 34.88±1.96         & \textbf{35.58±1.74} \\ \cline{3-8} 
                        &                       & SSIM   & 0.84±0.02   & 0.86±0.02  & 0.89±0.02  & 0.90±0.02          & \textbf{0.91±0.02}  \\ \cline{3-8} 
                        &                       & RMSE   & 10.30±11.00 & 1.55±1.20  & 1.49±1.27  & 1.37±0.93          & \textbf{1.14±0.66}  \\ \cline{2-8} 
                        & \multirow{3}{*}{6X}   & PSNR   & 25.15±3.37  & 32.67±1.89 & 30.83±2.09 & 32.34±2.16         & \textbf{33.36±1.81} \\ \cline{3-8} 
                        &                       & SSIM   & 0.79±0.03   & 0.82±0.03  & 0.84±0.02  & 0.86±0.02          & \textbf{0.87±0.02}  \\ \cline{3-8} 
                        &                       & RMSE   & 18.2±22.3   & 2.36±2.09  & 4.08±4.29  & 2.62±2.19          & \textbf{2.00±1.52}  \\ \hline
\multirow{9}{*}{Radial} & \multirow{3}{*}{50\%} & PSNR   & 35.17±1.37  & 36.70±1.37 & 37.17±1.37 & 38.38±1.22         & \textbf{38.91±1.21} \\ \cline{3-8} 
                        &                       & SSIM   & 0.90±0.02   & 0.91±0.01  & 0.92±0.01  & \textbf{0.93±0.01} & \textbf{0.93±0.01}  \\ \cline{3-8} 
                        &                       & RMSE   & 2.37±0.50   & 1.97±0.59  & 1.53±0.18  & 1.22±0.16          & \textbf{1.14±0.15}  \\ \cline{2-8} 
                        & \multirow{3}{*}{30\%} & PSNR   & 32.69±1.13  & 34.02±1.39 & 35.27±1.40 & 35.96±1.27         & \textbf{36.21±1.29} \\ \cline{3-8} 
                        &                       & SSIM   & 0.80±0.03   & 0.87±0.02  & 0.88±0.01  & 0.88±0.02          & \textbf{0.89±0.02}  \\ \cline{3-8} 
                        &                       & RMSE   & 4.87±0.89   & 2.23±0.76  & 1.39±0.67  & 1.61±0.22          & \textbf{1.36±0.32}  \\ \cline{2-8} 
                        & \multirow{3}{*}{20\%} & PSNR   & 30.77±1.15  & 33.69±1.51 & 33.49±1.54 & 34.48±1.26         & \textbf{34.72±1.30} \\ \cline{3-8} 
                        &                       & SSIM   & 0.75±0.03   & 0.84±0.02  & 0.85±0.02  & \textbf{0.86±0.02} & \textbf{0.86±0.02}  \\ \cline{3-8} 
                        &                       & RMSE   & 7.13±1.24   & 2.76±0.87  & 2.68±0.73  & 1.91±0.27          & \textbf{1.61±0.34}  \\ \hline
\multirow{9}{*}{Spiral} & \multirow{3}{*}{50\%} & PSNR   & 35.62±1.27  & 39.37±1.39 & 41.39±1.37 & 42.53±0.75         & \textbf{43.04±0.77} \\ \cline{3-8} 
                        &                       & SSIM   & 0.91±0.02   & 0.94±0.01  & 0.94±0.02  & 0.96±0.01          & \textbf{0.97±0.02}  \\ \cline{3-8} 
                        &                       & RMSE   & 1.78±0.54   & 1.06±0.10  & 0.97±0.14  & 0.73±0.15          & \textbf{0.61±0.07}  \\ \cline{2-8} 
                        & \multirow{3}{*}{30\%} & PSNR   & 33.48±1.12  & 36.19±1.35 & 36.26±1.18 & 38.20±1.31         & \textbf{38.49±1.33} \\ \cline{3-8} 
                        &                       & SSIM   & 0.86±0.02   & 0.90±0.01  & 0.90±0.02  & \textbf{0.92±0.01} & \textbf{0.92±0.01}  \\ \cline{3-8} 
                        &                       & RMSE   & 3.71±0.91   & 1.67±0.72  & 1.64±0.35  & 1.24±0.18          & \textbf{0.98±0.23}  \\ \cline{2-8} 
                        & \multirow{3}{*}{20\%} & PSNR   & 30.97±1.14  & 34.88±1.39 & 33.83±1.50 & 36.51±1.23         & \textbf{36.75±1.27} \\ \cline{3-8} 
                        &                       & SSIM   & 0.81±0.03   & 0.88±0.02  & 0.88±0.01  & \textbf{0.90±0.02} & \textbf{0.90±0.02}  \\ \cline{3-8} 
                        &                       & RMSE   & 5.39±1.26   & 2.29±0.56  & 1.95±0.25  & 1.51±0.20          & \textbf{1.32±0.14}  \\ \hline
\end{tabular}%
}
\label{table_knee}
\end{table}

The quantitative metrics (PSNR, SSIM and RMSE) of each GAN based method using different under-sampling masks are shown in Tables \ref{table_brain} and \ref{table_knee}. We can draw similar conclusions as the qualitative visualisation results that the image quality after RefineGAN reconstruction is better than other methods. Even at a high acceleration factor or high under-sampling rate, the reconstructed images of RefineGAN still have high SNR.

Tables \ref{table_brain} and \ref{table_knee} show the quantitative metrics, including PSNR, SSIM and RMSE for all compared methods. The numbers in Tables \ref{table_brain} and \ref{table_knee} represent the mean and standard deviation values of the corresponding metrics (bold numbers indicate the best performance). Compared to DAGAN, KIGAN and ReconGAN, the RefineGAN framework has outperformed them remarkably at different acceleration factors.

\section{Discussion}
We have established that GAN based methods such as DAGAN, KIGAN, ReconGAN, and RefineGAN excel in generating faithful, photo-realistic reconstruction of undersampled MR images and in removing the undersampling artefacts.  Despite their representing a successful category of CS-MRI techniques, GAN based methods suffer from training instability and slow convergence to a global minimum \cite{Chen2017,Wiatrak2020}. This problem can be alleviated by Wasserstein GAN (WGAN) \cite{Arjovsky2017}.  Instead of minimising Jensen-Shannon divergence between the reconstructed image and ground truth in the training phase, WGAN aims to reduce the Wasserstein distance.  WGAN based CS-MRI models have achieved superior reconstruction performance compared with GAN based methods e.g., DAGAN \cite{jiang_accelerating_2019,oh_unpaired_2020}. However, to minimise the Wasserstein distance, the discriminator needs to satisfy the 1-Lipschitz constraint. The original WGAN paper adopted a weight clipping approach to enforce this constraint but this can itself cause training instability. The potential improvement includes a gradient clipping approach \cite{Gulrajani2017} or dividing the weights of neural networks by their spectral norm \cite{miyato_spectral_2018}. Altogether, modifying the loss function of GAN is key to addressing its training instability issue.

Another problem with the GAN loss function is that it may affect the high frequency texture \cite{Mardani2019} and smoothing the reconstructed image \cite{zhao_loss_2017}. GAN may also generate high frequency noise \cite{Mardani2019}. Mardani et al. (2019) suggested solving the high frequency texture issue with the least square GAN \cite{Mardani2019}. L1 loss between reconstructed images and ground truth was also added as an auxiliary function, to serve as a low-pass filter to remove high frequency noises.

Apart from the issues with the GAN loss function during training, GAN based methods suffer from reconstruction instability \cite{antun_instabilities_2020}. This means that a small perturbation to the input image leads to large scale changes in the reconstruction output.  Antun et al. (2020) \cite{antun_instabilities_2020} evaluated the instabilities of DAGAN and other deep learning based CS-MRI models and showed that the performance of DAGAN deteriorated as the undersampling ratio increased during testing. That is if more \textit{k}-space pixels were undersampled more than the undersampling ratio in the images used to train DAGAN, its performance decreases. In contrast, variational network and deep cascaded convolutional neural networks (DC-CNN), two non-GAN based techniques, did not experience the same issue.  This indicates the incorporation of GAN may reduce the generalisability of the model to images collected under different undersampling ratios and contribute to its reconstruction instability.

Another instability issue with DAGAN is that if small structural details were added to an MRI image, such as random letters, and this image was undersampled retrospectively, DAGAN was unable to reconstruct the structural details. In contrast, variational network and DC-CNN were superior in recovering these structural details \cite{antun_instabilities_2020}. Even though these two instability tests revealed a critical weakness of DAGAN, DAGAN, being one of the earliest GAN based methods, may not capture the improved reconstruction accuracy and network architectural complexity of the more recent methods \cite{jiang_accelerating_2019,oh_unpaired_2020}. Hence, more instability testing is required on more GAN based methods to conclude whether this category of CS-MRI techniques suffers from reconstruction instability.

These results on instability testing still question the generalisability of GAN based models. In other words, it is unclear whether after being trained with images from one  source, e.g., one organ, under one undersampling ratio etc, a model will achieve equivalent performance on images from another source. This is often referred to in the literature as a zero-shot inference test. However, GAN based methods appear to excel in these inference tests. To illustrate, DAGAN, having been trained with MR images of the brain of a healthy volunteer, successfully reconstructed the MR images from a patient with the brain tumour, faithfully preserving the tumour structure \cite{Yang2018}. Another example is that GANCS, another GAN based method, was trained with MR images of the abdomen and any images containing the enlarged adrenal gland were removed from the training set. In testing, GANCS was capable of recovering enlarged adrenal gland in patients with adrenal hypertrophy. No evidence of hallucination of the reconstructed images was found \cite{Mardani2019}. These successes in the zero-shot inference of pathological structures challenged the instability test results from Antun et al. One interpretation is that the additional structural details added to the MR images by Antun et al., such as letters, may not represent real-life physiological or pathological deviations of the MR images from the training set. It was also possible that the zero-shot inference tests on DAGAN and GANCS concerned only a small sample size, limited to a few pathological conditions. Larger scale zero-shot inference tests are necessary to evaluate the generalisability of GAN based CS-MRI methods.

In summary, future research directions on GAN based fast MRI may include more robust training strategies, e.g., combining GAN with genetic algorithm \cite{hao2020annealing}, incorporating edge and texture enhancement \cite{chen2021wavelet}, reducing possible hallucination \cite{zhu2019can,zhu2019lesion}, coupling with explainable AI (XAI) modules \cite{ye2021explainable,yang2021unbox}, and further consideration with MR physics \cite{lv2021pic}.

\section{Conclusion}

We carried out a mini review, benchmarked and compared four different GAN-based network architectures for fast MRI reconstruction in this chapter. Our comparison used the various sampling patterns, different masks on corresponding datasets that have covered commonly used clinical MRI scenarios. For our systematic research and measurement, we used both traditional and newly proposed quantitative tools. The outcomes of qualitative visualization were also examined and compared. To summarise, our mini review and benchmarking have revealed that all GAN-based approaches could obtain promising results at lower acceleration factors. However, when the acceleration factors are high, some GAN-based architectures, such as DAGAN and KIGAN, may not be sufficient for MRI reconstruction applications. Furthermore, as compared to other GAN-based methods, the RefineGAN has improved reconstruction accuracy and perceptual efficiency. Future development incorporating MR physics and XAI into GAN based models will provide promising pathways for clinical deployment with more transparency of the reconstruction algorithms.

\newpage
\bibliography{GAN_MRI} 
\bibliographystyle{unsrt} 

\end{document}